\documentclass[twocolumn,aps,pra,superscriptaddress,showpacs]{revtex4-1} 
\usepackage[english]{babel}
\usepackage[utf8]{inputenc}
\usepackage{braket}
\usepackage[colorinlistoftodos, color=green!40, prependcaption]{todonotes}
\usepackage{amsthm}
\usepackage{mathtools}
\usepackage{physics}
\usepackage{xcolor}
\usepackage{graphicx}
\usepackage[left=23mm,right=13mm,top=35mm,columnsep=15pt]{geometry} 
\usepackage{adjustbox}
\usepackage{placeins}
\usepackage[T1]{fontenc}
\usepackage{lipsum}
\usepackage{csquotes}
\usepackage[pdftex, pdftitle={Article}, pdfauthor={Author}]{hyperref} 
\newcommand{\cm}{cm$^{-1}$}
\usepackage{multirow}
\usepackage{threeparttable}
\usepackage{fancyhdr} 
\usepackage{color,soul}
\usepackage{afterpage}

\begin{document}
\title{Extending the large molecule limit: The role of Fermi resonance in developing a quantum functional group}

\author{Guo-Zhu Zhu}
\affiliation{Department of Physics and Astronomy, University of California, Los Angeles, California 90095, USA}

\author{Guanming Lao}
\affiliation{Department of Physics and Astronomy, University of California, Los Angeles, California 90095, USA}

\author{Claire E. Dickerson}
\affiliation{Department of Chemistry and Biochemistry, University of California, Los Angeles, California 90095, USA}

\author{Justin R. Caram}
\affiliation{Department of Chemistry and Biochemistry, University of California, Los Angeles, California 90095, USA}
\affiliation{Center for Quantum Science and Engineering, University of California, Los Angeles, California 90095, USA}

\author{Wesley C. Campbell}
\affiliation{Department of Physics and Astronomy, University of California, Los Angeles, California 90095, USA}
\affiliation{Center for Quantum Science and Engineering, University of California, Los Angeles, California 90095, USA}
\affiliation{Challenge Institute for Quantum Computation, University of California, Los Angeles, California 90095, USA}

\author{Anastassia N. Alexandrova}
\affiliation{Department of Chemistry and Biochemistry, University of California, Los Angeles, California 90095, USA}
\affiliation{Center for Quantum Science and Engineering, University of California, Los Angeles, California 90095, USA}

\author{Eric R. Hudson}
\affiliation{Department of Physics and Astronomy, University of California, Los Angeles, California 90095, USA}
\affiliation{Center for Quantum Science and Engineering, University of California, Los Angeles, California 90095, USA}
\affiliation{Challenge Institute for Quantum Computation, University of California, Los Angeles, California 90095, USA}

\date{\today} 

\begin{abstract}
Polyatomic molecules equipped with optical cycling centers (OCCs), enabling continuous photon scattering during optical excitation, are exciting candidates for advancing quantum information science. 
However, as these molecules grow in size and complexity the interplay of complex vibronic couplings on optical cycling becomes a critical, but relatively unexplored consideration. 
Here, we present an extensive exploration of Fermi resonances in large OCC-containing molecules, surpassing the constraints of harmonic approximation. 
High-resolution dispersed laser-induced fluorescence and excitation spectroscopy reveal Fermi resonances in calcium and strontium phenoxides and their derivatives. 
This resonance manifests as vibrational coupling leading to intensity borrowing by combination bands near optically active harmonic bands.
The resulting additional vibration-changing decays require more repumping lasers for effective optical cycling. 
To mitigate these effects, we explore altering vibrational energy level spacing through substitutions on the phenyl ring or changes in the OCC itself.
While the complete elimination of vibrational coupling in complex molecules remains challenging, our findings underscore the potential for significant mitigation, opening new avenues for optimizing optical cycling in large polyatomic molecules.
\end{abstract}

\maketitle

Functionalizing large molecules with optical cycling centers (OCCs) is being explored as a means for extending the exquisite control available in quantum information science to the chemical domain.\cite{Isaev2016Polyatomic,kozyryev2016proposal,ivanov2019towards,ivanov2020two,ivanov2020toward,klos2020prospects,augenbraun2020molecular,Dickerson2021FranckCondon,dickerson2021optical,liu2021electronic,liu2022combined,dickerson2022fully,yu2023multivalent,sinenka2023zwitterions,zhu2022caoph}
Success requires that these OCCs absorb and emit many photons without changing vibrational states.
To accomplish this task, molecular design rules are being developed, aided and validated by experiments, to guide the creation of the ideal quantum functional groups~\cite{zhu2022caoph,mitra2022pathway,lao2022sroph,changala2023structural}.
For example, prior work has demonstrated that alkaline earth alkoxides 
provide a general and versatile chemical moiety for optical cycling applications, as the alkaline earth radical electron can be excited without perturbing the vibrational structure of the molecule~\cite{liu2021electronic,liu2022combined,zhu2022caoph,mitra2022pathway,lao2022sroph}.
Similarly, traditional physical organic principles, such as electron-withdrawing, have been shown to improve OCCs performance~\cite{zhu2022caoph,Dickerson2021FranckCondon}.
Further, experimental and theoretical extensions to more complex acenes, \cite{dickerson2021optical,mitra2022pathway} diamondoid \cite{dickerson2022fully} and even surfaces \cite{guo2021surface} suggest an exciting path forward for creating increasingly complex and functional quantum systems.
 
However, an open question for this work is what role intramolecular vibrational energy redistribution (IVR) will play as the molecule size is further increased~\cite{uzer1991ivr,nesbitt1996IVR,keske2000decoding}. 
In the typical description of IVR, the normal modes of molecular vibrations are treated within the harmonic approximation, while any anharmonic couplings between these modes are treated as a perturbation. 
Laser excitation to an excited (harmonic) vibrational state is then followed by the redistribution of the vibrational energy driven by the anharmonic couplings. 
This outflow of energy from one vibrational mode to other modes is an artifact of the choice of basis states that are not eigenstates of the molecular Hamiltonian, and thus not stationary.

An alternate, and equivalent, description of IVR takes the vibrational eigenstates of the molecular Hamiltonian as the basis.
These basis states are mixtures of the harmonic vibrational modes, with amplitudes set by the anharmonic couplings. 
As these states are eigenstates of the molecular Hamiltonian they are, of course, not time-evolving (except for their coupling to the electromagnetic vacuum) and therefore there is no energy redistribution between them unless perturbed by an external field or collision. 
Instead, the effect of IVR in this picture is simply that there is more than one vibronic state within the spectrum of the exciting laser leading to non-exponential fluorescence as decay from these nearby states interfere. 

This latter picture is convenient for understanding the role that IVR will play in functionalizing large molecules with OCCs. 
If harmonic vibrational states are close together and possess the correct symmetry, then anharmonic couplings will mix them.
In this case, a harmonic vibrational state, which is not optically active, becomes optically active by mixing with an optically active harmonic mode. 
While this does not change the fraction of vibration-changing decays,
it does change the number of accessible final vibrational states and requires more repumping lasers to achieve optical cycling~\cite{mccarron2018laser,fitch2021laser,augenbraun2023direct}.

Therefore, to push optical cycling to larger and larger molecules it is necessary to develop molecular design principles for avoiding these vibrational couplings by energy separation and/or symmetry. 
Here, we explore these phenomena in both the calcium and strontium phenoxides, which have recently been shown as promising candidates for optical cycling~\cite{zhu2022caoph,lao2022sroph,augenbraun2022caoph}. 
We show that in certain derivatives of these molecules it is possible to find combination modes (within the harmonic approximation), which are not themselves optically active, close to optically-active stretching modes. 
Anharmonic coupling between these modes -- e.g. Fermi resonance,\cite{fermi1931ramaneffekt,dubal1984tridiagonal} 
which is the simplest instance of IVR -- leads to intensity borrowing and the activation of the combination mode so that a new decay pathway is opened.
Such molecules will require extra repumping lasers for optical cycling. 
By comparing phenoxides with and without this effect, we present further design rules for functionalizing ever larger molecules with optical cycling centers.

\begin{figure}
    \centering
    \includegraphics[scale=0.4]{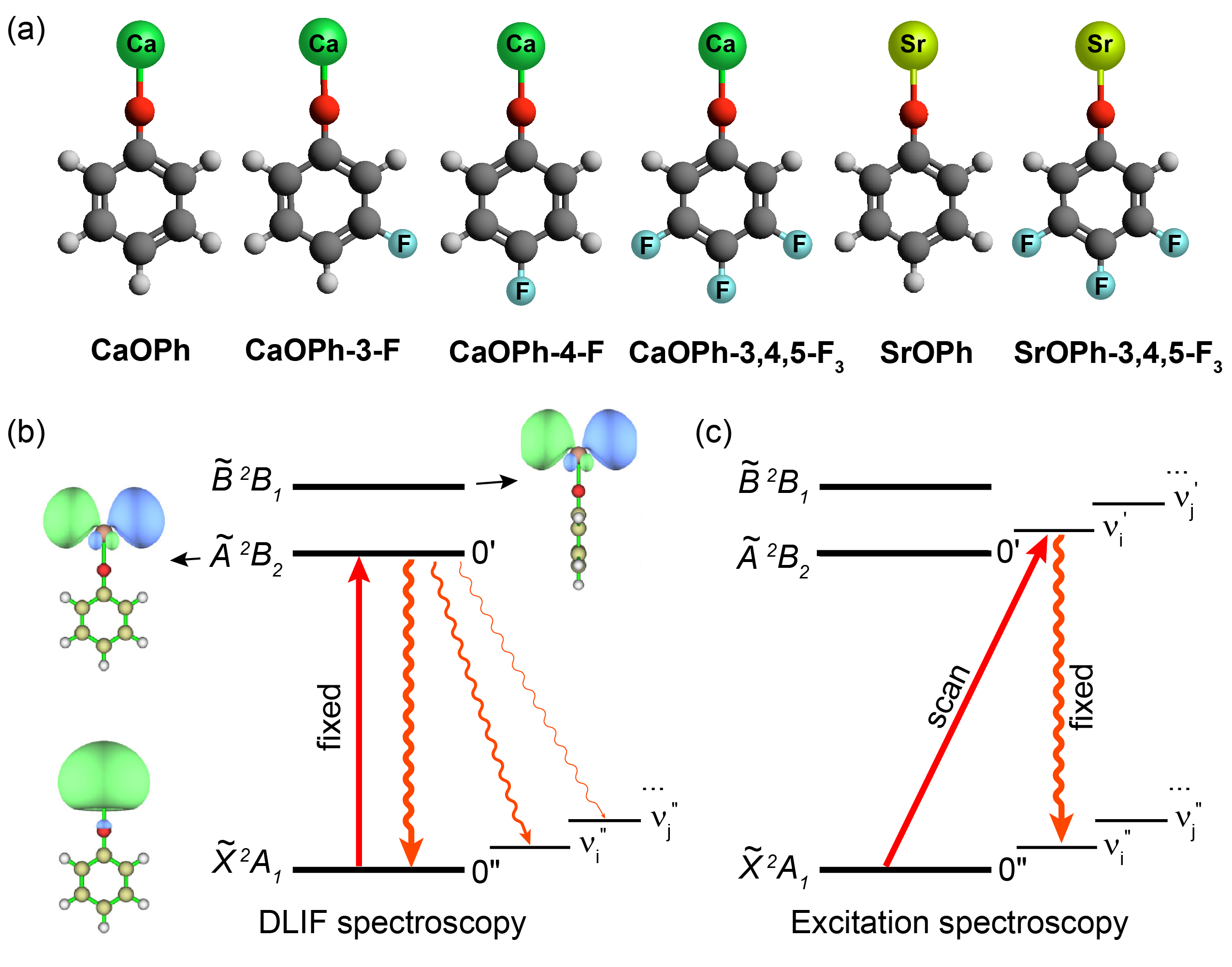}
    \caption{(a) Molecular structures of all studied calcium and strontium phenoxide and derivatives. (b) and (c) Schematic diagrams illustrating the DLIF measurement and excitation spectroscopy performed for all molecules in this study, respectively. DLIF measurements were done by fixing the laser wavelength at the transitions of $\widetilde A/\widetilde B (v'=0) \leftarrow \widetilde X (v''=0)$ and detecting the dispersed fluorescence. Excitation spectroscopy was carried out with off-diagonal excitation scan, $\widetilde A/\widetilde B (v_n') \leftarrow \widetilde X (v''=0)$, and diagonal fluorescence detection $\widetilde A/\widetilde B (v_n') \rightarrow \widetilde X (v_n'')$. The molecular orbital and symmetries of the electronic states are based on the CaOPh molecule with a C$_{2v}$ symmetry.
    } 
    \label{fig:molecule} 
\end{figure}

A series of calcium and strontium phenoxides (CaOPh, CaOPh-3-F, CaOPh-4-F, CaOPh-3,4,5-F$_3$, SrOPh, and SrOPh-3,4,5-F$_3$, Ph = phenyl group, Fig. \ref{fig:molecule}a) were produced via laser ablation of the alkaline earth metal into a mixture of the precursor ligand and Ne buffer gas inside cryogenic cell operated at a temperature range of $\sim$20~K (Fig. \ref{fig:setup}) \cite{lao2022sroph}.
As sketched in Figs. \ref{fig:molecule}b-c, the vibrational structure of these molecules was probed with two types of measurements: dispersed laser-induced fluorescence (DLIF) spectroscopy, which probes the vibrational structure in the electronic ground state ($\widetilde X$), and excitation spectroscopy, which examines the vibrational structure in the excited states ($\widetilde A$ and $\widetilde B$). 
In DLIF spectroscopy (Fig.~\ref{fig:molecule}b), vibrationally cold molecules are excited to the ground vibrational level of the electronically excited $\widetilde A$ and $\widetilde B$ states, $\widetilde A/\widetilde B (v'=0) \leftarrow \widetilde X (v''=0)$, and the resulting fluorescence is recorded as a function of wavelength.
In excitation spectroscopy (Fig.~\ref{fig:molecule}c), the exciting laser is tuned to drive excitation to excited vibrational levels of the excited $\widetilde A$ and $\widetilde B$ states, $\widetilde A/\widetilde B (v_n') \leftarrow \widetilde X (v''=0)$, while simultaneously monitoring the resulting fluorescence from non-vibration-changing decays, $\widetilde A/\widetilde B (v_n') \rightarrow \widetilde X (v_n'')$.
In both cases, excitation is provided via a tunable pulsed dye laser and the resulting fluorescence is coupled into a grating monochromator and detected using a photomultiplier tube. 
Compared to previous measurements \cite{zhu2022caoph,lao2022sroph}, improvements, such as better source handling techniques to reduce the production of alkaline earth oxide contaminants, provided an increase in signal-to-noise ratio (SNR) of $\sim$3$\times$. 
This improved SNR enabled spectrometer measurements with a higher resolution of 0.20 nm. 
Additional experimental details and theoretical methods are provided in the Supplementary Information.

\begin{figure*}
    \centering
    \includegraphics[scale=0.45]{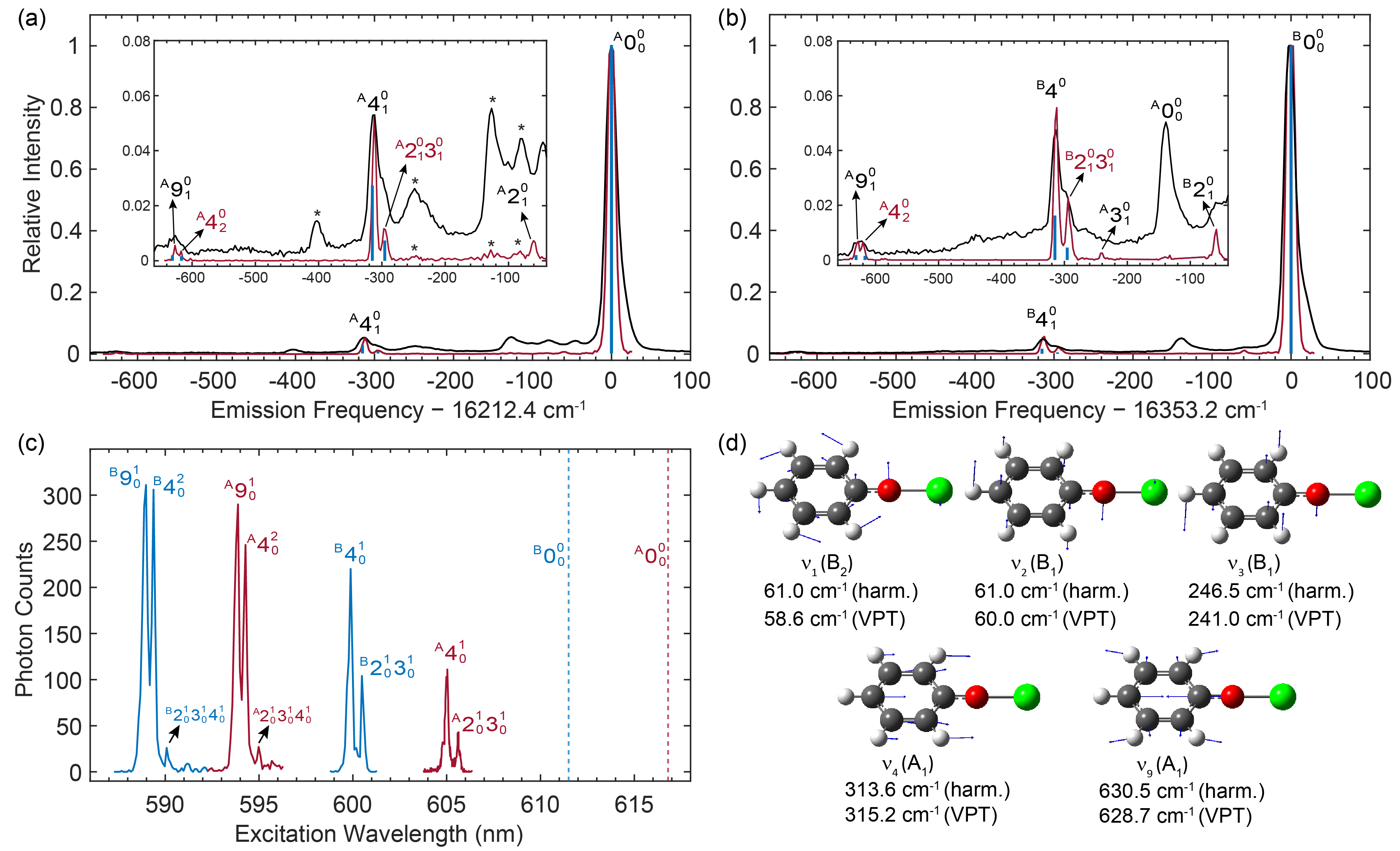}
    \caption{(a) and (b) Dispersed fluorescence spectra obtained for the $\widetilde A \rightarrow \widetilde X$ and $\widetilde B \rightarrow \widetilde X$ transitions of CaOPh. The black traces, adopted from Ref.\cite{zhu2022caoph}, were obtained with a spectral resolution of $\approx$ 0.5~nm. The red traces come from an improved measurement with resolution of $\approx$ 0.20 nm. The insets display the weak peaks in the range of $-660$ \cm~to $-40$ \cm~and show the presence of doublet peaks around $-300$ \cm~and $-630$ \cm. 
    The blue sticks depict the calculated frequencies and relative strengths (FCFs) of the vibrational modes using the VPT method. The symbol * indicates CaOH contamination. (c) Excitation spectrum of the $\widetilde A (v_n') \leftarrow \widetilde X (v'' = 0)$ (red traces) and $\widetilde B (v_n') \leftarrow \widetilde X (v'' = 0)$ (blue traces) transitions. The resulting fluorescence is monitored on decays that do not change the vibrational quantum number. The two dashed lines indicate the excitation wavelengths corresponding to the respective 0-0 transitions. The assignments of all observed vibrational resonances are given. 
    (d) Vibrational displacements of five related fundamental modes. The symmetries and theoretical frequencies in $\widetilde X$ using harmonic and VPT methods are provided. 
    } 
    \label{fig:CaOPh_DLIF}
\end{figure*}

\begin{figure*}
    \centering
    \includegraphics[scale=0.45]{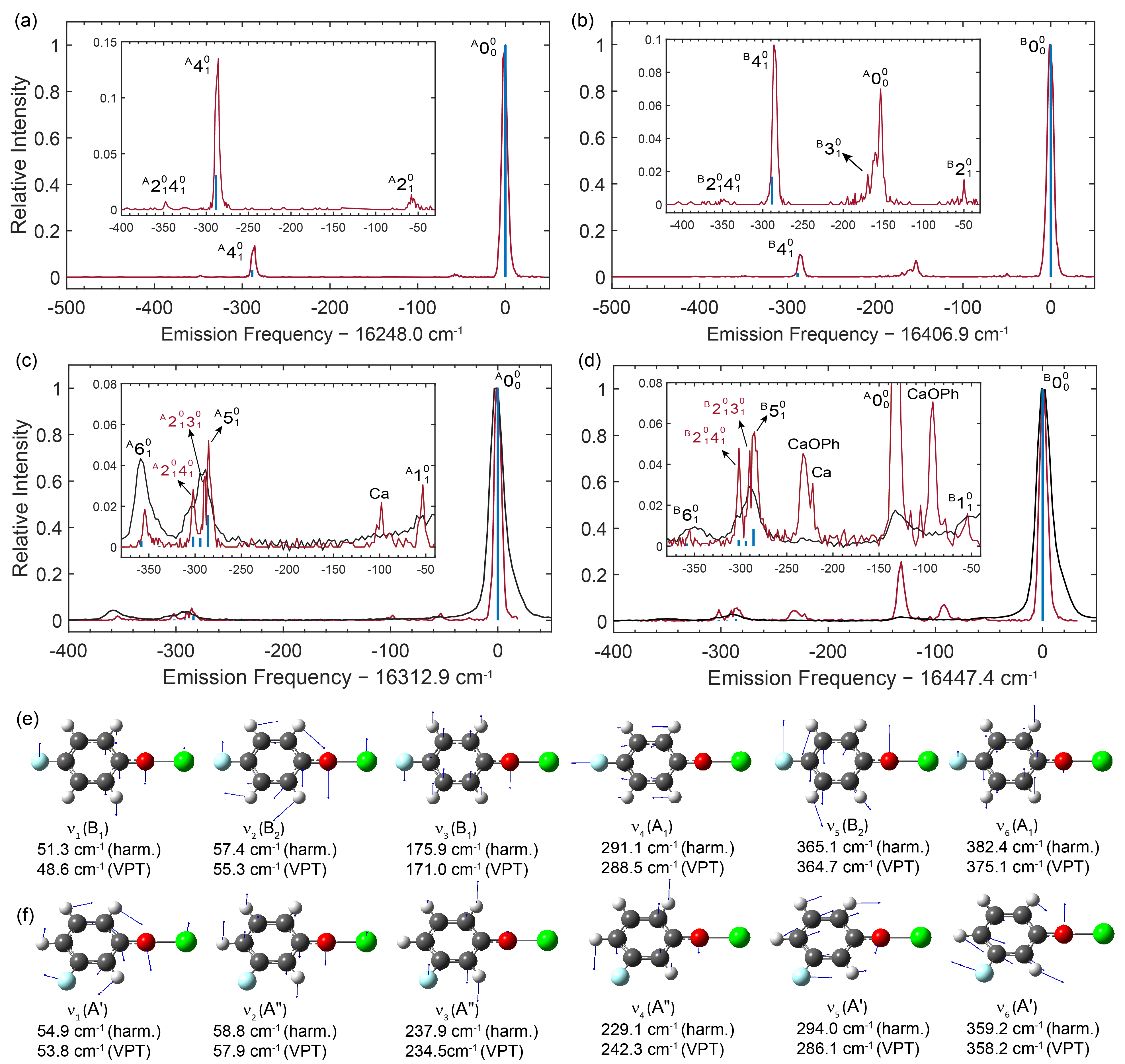}
    \caption{(a) and (b) Dispersed spectra for the $\widetilde A \rightarrow \widetilde X$ and $\widetilde B \rightarrow \widetilde X$ transitions of CaOPh-4-F molecule, respectively. Only a single peak is observed for the stretching mode $\nu_4$ around $-$286 \cm. Due to the absence of Fermi resonance coupling, the theoretical relative strengths ( blue vertical lines) are calculated under harmonic approximation. (c) and (d) Dispersed spectra for the $\widetilde A \rightarrow \widetilde X$ and $\widetilde B \rightarrow \widetilde X$ transitions of CaOPh-3-F molecule, respectively. The black traces are taken from ref.\cite{zhu2022caoph}, measured with a spectral resolution of $\approx$ 0.5 nm, while the red traces represent an improved measurement with resolution of $\approx$ 0.20 nm. Three decays near $-290$~\cm~ are observed. The blue vertical lines indicate the calculated vibrational frequencies and relative strengths using the VPT method. (e) and (f) Vibrational displacements of the six lowest-frequency fundamental modes in the ground state. Theoretical frequencies and symmetries for these modes are given. 
    } 
    \label{fig:3F&4F_DLIF}
\end{figure*}

Using this improved resolution, we recorded DLIF spectra for the $\widetilde A \rightarrow \widetilde X$ and $\widetilde B \rightarrow \widetilde X$ transitions of CaOPh, as shown as the red lines in 
Figs. \ref{fig:CaOPh_DLIF}a and \ref{fig:CaOPh_DLIF}b, respectively.
For comparison, the previously recorded DLIF spectra for this molecule from Ref.~\cite{zhu2022caoph} are shown as black lines. 
Several improvements are immediately clear.
First, spectral contamination by CaOH molecules, features denoted by *, is greatly reduced.
Second, while 
in the previous work three fundamental vibrational modes ($\nu_2$, $\nu_4$ and $\nu_9$) were resolved within the frequency range of $\sim$660~\cm~below the respective 0-0 transition, the improved measurements here reveal several new features which were either unresolved in or below the detection limit of the previous measurement. Specifically, the lowest-frequency out-of-plane bending mode $\nu_2$ (Fig. \ref{fig:CaOPh_DLIF}d) is much better resolved at a frequency shift of $-60$ \cm~(Figs. \ref{fig:CaOPh_DLIF}a-b). A new weak decay is also observed at $-241$ \cm~ (Fig. \ref{fig:CaOPh_DLIF}b) and readily assigned to the fundamental out-of-plane bending mode $\nu_3$ (Fig. \ref{fig:CaOPh_DLIF}d).
Further, the previously assigned peaks due to decay to the stretching modes $\nu_4$ and $\nu_9$ are seen to be doublets. While theoretical calculations within the harmonic approximation predict $\nu_4$ should be the strongest vibration-changing decay and occur at $-313.6$~\cm~(Table \ref{tab:DLIF-vbr}), the weaker peak at $-295$~\cm~is not readily assignable. Compared with the theoretical harmonic vibrational frequencies, the weak peak is near the combination modes, $\nu_1\nu_3$ (theo., 307.5~\cm) or $\nu_2\nu_3$ (theo., 307.5~\cm), as shown in Fig. \ref{fig:CaOPh_DLIF}d, however, the predicted Franck-Condon factors (FCFs) for these decays is <10$^{-4}$, well below the current detection limit. 
The observed decay can be explained by an intensity borrowing mechanism \cite{zhang2021accurate,zhang2023intensity}, which arises from anharmonic coupling between the nearly degenerate stretching mode $\nu_4$ and the combination mode consisting of two bending modes, also known as a Fermi resonance~\cite{fermi1931ramaneffekt,dubal1984tridiagonal,nesbitt1996IVR}.  
To corroborate Fermi resonance splittings, vibrational perturbation theory (VPT) with resonances was applied on top of anharmonic frequency calculations to predict corrected frequencies, resonance splittings, and obtain anharmonic FCFs (see Supplementary Information for more details). 

As seen in the insets of Figs. \ref{fig:CaOPh_DLIF}a-b, the predicted splittings (vertical blue lines) agree well with the observed vibrational doublets (red traces).
Given the requirement that coupled vibrational modes have the same symmetry, the weak peak is attributed to the $\nu_2\nu_3$ combination mode with A$_1$ symmetry, rather than $\nu_1\nu_3$ with A$_2$ symmetry (Fig. \ref{fig:CaOPh_DLIF}d). 
Similarly, the doublet near $\nu_9$ is interpreted as a result of vibrational decays to a fundamental mode $\nu_9$, as observed previously~\cite{zhu2022caoph}, and the overtone of the stretching mode $\nu_4$. Within the harmonic approximation (Table \ref{tab:DLIF-vbr}), the decay intensity of $\nu_9$ is expected to be approximately four times that of the overtone of mode $\nu_4$ for the $\widetilde A \rightarrow \widetilde X$ (and ten times for the $\widetilde B \rightarrow \widetilde X$ transition). 
The observed nearly equal intensities in both transitions in Figs. \ref{fig:CaOPh_DLIF}a-b are due to the intensity borrowing via Fermi resonance.

The presence of vibrational doublets due to anharmonic couplings is also observed in the electronically excited $\widetilde A$ and $\widetilde B$ states by excitation spectroscopy, as presented in Fig. \ref{fig:CaOPh_DLIF}c. 
Here, it is seen that for both electronically excited states, as in the ground state, the Fermi resonance leads to activation of the $\nu_2\nu_3$ combination mode at a splitting of around 16~\cm~from the $\nu_4$ vibrational level (Table \ref{tab:vib-in-AB-Ca}).  
Similarly, excitations to the excited vibrational levels of $\nu_9$ and 2$\nu_4$, as well as a very weak resonance to the combination band of $\nu_2\nu_3\nu_4$, are observed. 
The observation of the vibrational anharmonic coupling across different electronic states highlights the significance of Fermi resonances in the spectral characteristics of large molecules like CaOPh. 

To explore the universality of Fermi resonances, we extended our study to the substituted molecules CaOPh-4-F, CaOPh-3-F and CaOPh-3,4,5-F$_3$. 
In Figs. \ref{fig:3F&4F_DLIF} and \ref{fig:345F-DLIF}, the DLIF spectra of the $\widetilde A \rightarrow \widetilde X$ and $\widetilde B \rightarrow \widetilde X$ transitions for these substituted molecules are presented. 
Remarkably, with a single fluorine atom substituted at the \text{para}-position of the phenyl ring, the DLIF spectra of CaOPh-4F (Figs. \ref{fig:3F&4F_DLIF}a-b) show only a single peak for the vibrational decay to the stretching mode $\nu_4$ for both transitions. 
This implies the absence of a Fermi resonance, which can be attributed to the substantial frequency splitting of 64 \cm~(harm.) or 69 \cm~(VPT) between $\nu_4$ 
and the symmetry-allowed combination band of $\nu_1\nu_3$ (Fig. \ref{fig:3F&4F_DLIF}e). 
Furthermore, the insets in Figs. \ref{fig:3F&4F_DLIF}a-b reveal two weak peaks at frequencies of around $-$53 \cm~and $-$346 \cm, which can be assigned to mode $\nu_2$ and $\nu_2\nu_4$, respectively, by comparing with theoretical frequencies (Fig. \ref{fig:3F&4F_DLIF}e). 
These weak peaks are likely due to the anharmonic mode-coupling involving the low-frequency bending mode $\nu_2$ \cite{zhu2022caoph}. 
Additionally, the complex peaks observed at around $-$150~\cm~result from collision-induced relaxation from $\widetilde B \rightarrow \widetilde A$, followed by fluorescence decay to the $\widetilde X$ state, and a vibrational decay to mode $\nu_3$ at $-$170~\cm.

In the case of CaOPh-3-F, where the \textit{para}-F is replaced with a \textit{meta}-F and the molecular symmetry is reduced from C$_{2v}$ to C$_s$, the coupling phenomenon is markedly different. 
While previous DLIF studies~\cite{zhu2022caoph} of $\widetilde A \rightarrow \widetilde X$ and $\widetilde B \rightarrow \widetilde X$ transitions found a broad peak for the stretching mode peak $\nu_5$ at $-$290 \cm (black traces in Figs. \ref{fig:3F&4F_DLIF}c-d), the present, higher resolution spectra, resolve three separate transitions, which are also predicted by the VPT calculation (blue lines in Figs. \ref{fig:3F&4F_DLIF}c-d).
The strongest peak at $-$284 \cm~corresponds to the vibrational decay to the stretching mode $\nu_5$ ($A'$, Fig. \ref{fig:3F&4F_DLIF}f), while the other two peaks at $-$291 \cm~and $-$302 \cm~are assigned to two combination bands, $\nu_2\nu_3$ ($A'$) and $\nu_2\nu_4$ ($A'$), respectively. 
This more complex coupling behavior can be attributed to the lower C$_s$ symmetry of CaOPh-3-F molecule. 
All three vibrational modes, $\nu_2$, $\nu_3$ and $\nu_4$, are out-of-plane bending modes with $A''$ symmetry. The combination of $\nu_2\nu_3$ or $\nu_2\nu_4$ results in $A'$ symmetry and frequencies close to that of the stretching mode $\nu_5$ (Fig. \ref{fig:3F&4F_DLIF}f), leading to intensity borrowing and activation of these unexpected combination bands.

\begin{figure}
    \centering
    \includegraphics[scale=0.45]{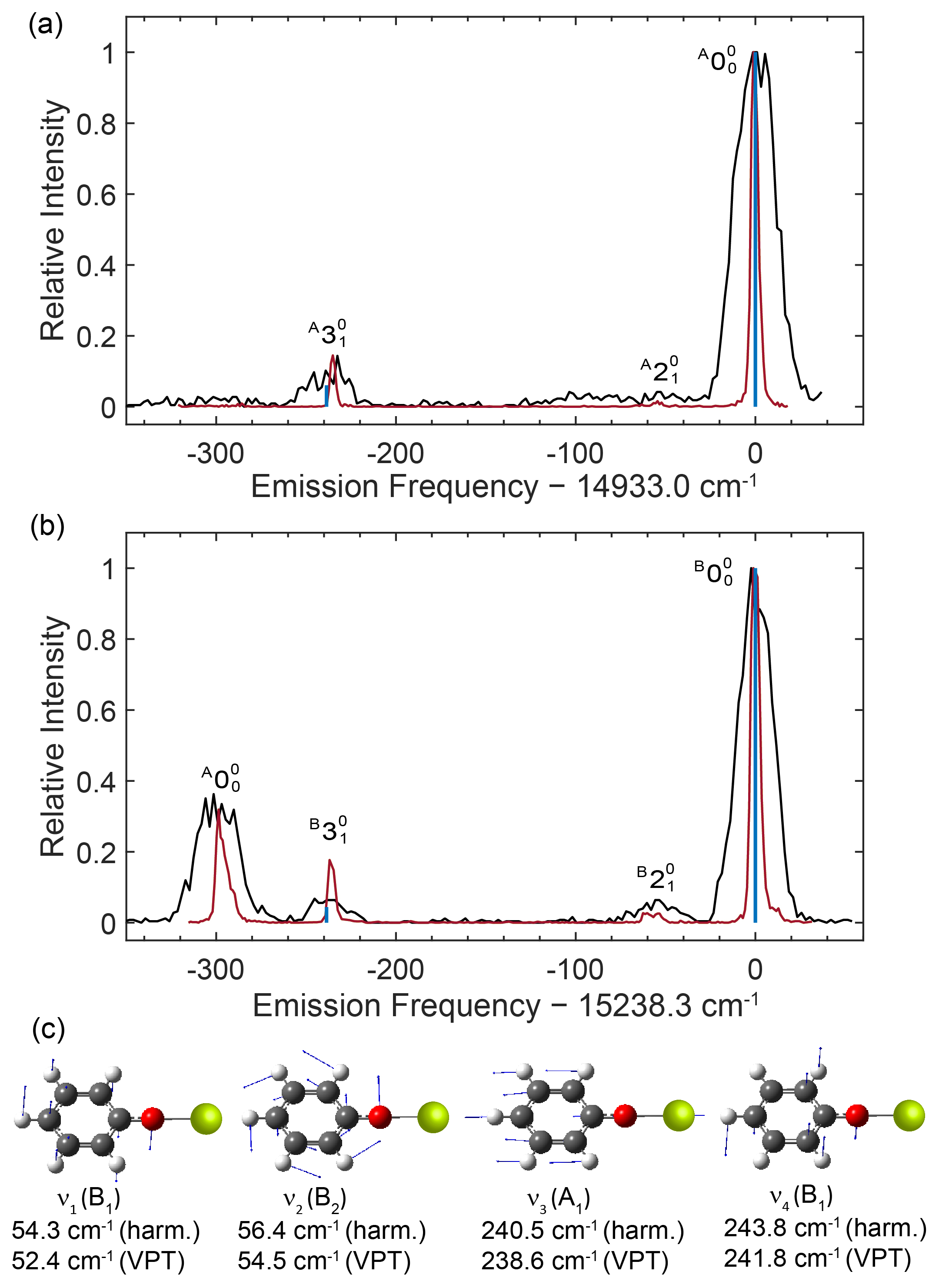}
    \caption{(a)-(b) DLIF spectra obtained for the $\widetilde A - \widetilde X$ and $\widetilde B - \widetilde X$ 
    transitions of SrOPh molecules. The black traces are taken from Ref.\cite{lao2022sroph}, measured with a spectral resolution of $\approx$ 0.5 nm, while the red traces represent an improved measurement with a resolution of $\approx$ 0.20 nm. The blue sticks show the calculated frequency (VPT) and relative strength (harm.) of vibrational decays. (c) Vibrational displacements of four lowest-frequency fundamental modes. Theoretical frequencies and symmetries for these modes are provided.} 
    \label{fig:SrOPh-DLIF}
\end{figure}

The absence of Fermi resonance in the CaOPh-4-F stretching mode decay and the presence of complex coupling in CaOPh-3-F are further supported by the excitation spectra obtained for the excited states. 
Fig. \ref{fig:4F-excitaiton} demonstrates a single peak corresponding to the stretching mode $\nu_4$ in the excitation spectra of CaOPh-4-F, while the excitation spectra of CaOPh-3-F (Fig. \ref{fig:3F-excitation}) reveal the presence of three transitions in the frequency region associated with the stretching mode $\nu_5$.

A more complex molecule with three F atoms substituted, CaOPh-3,4,5-F$_3$, has also been revisited as it is potentially the most attractive calcium phenoxide for optical cycling~\cite{zhu2022caoph}.
The DLIF spectra in Fig. \ref{fig:345F-DLIF} and excitation spectra of excited states in Fig. \ref{fig:345F-excitation} both reveal the presence of doublet vibrational peaks near the stretching mode peak region. 
One of these peaks corresponds to the stretching mode $\nu_6$ with an A$_1$ symmetry, while the other peak arises from a combination band involving two out-of-plane bending modes $\nu_1$ (B$_1$) and $\nu_4$ (B$_1$). 

To investigate the influence of metal atoms on anharmonic vibrational coupling, we have also studied two strontium phenoxides, SrOPh and SrOPh-3,4,5-F$_3$. 
Previous study~\cite{lao2022sroph} has provided low-resolution DLIF spectra for these molecules. 
Figs. \ref{fig:SrOPh-DLIF}a-b display the higher resolution DLIF spectra recorded here for SrOPh from the excited $\widetilde A$ and $\widetilde B$ states. 
Only a single transition is observed for the stretching mode $\nu_3$ at around $-$235 \cm, indicating the lack of Fermi resonance. 
The absence can be explained by the different symmetry of the combination band of $\nu_1\nu_2$ (A$_2$) and the stretching mode $\nu_3$ (A$_1$), along with a substantial energy gap of either 130~\cm~(harm.) or 132 \cm~(VPT), as shown in Fig. \ref{fig:SrOPh-DLIF}c. 
This is also validated by the presence of a single stretching mode transition in the excitation spectra of $\widetilde A \leftarrow \widetilde X$ and $\widetilde B \leftarrow \widetilde X$ in Fig. \ref{fig:sroph-excitation}.

Contrary to SrOPh, both DLIF spectra (Fig. \ref{fig:sroph345f-dlif}) and excitation spectra (Fig. \ref{fig:sroph345f-excitation}) of SrOPh-3,4,5-F$_3$ exhibit a weak transition assigned to the $\nu_1\nu_3$ mode in close proximity to the stretching-mode peak $\nu_4$, implying the existence of a small anharmonic coupling, as also captured by the VPT calculation.

The branching ratios and frequencies of all observed vibrational modes in the DLIF and excitation spectra are summarized in Table \ref{tab:DLIF-vbr}-\ref{tab:vib-in-AB-Sr}. 
From these, a consistent understanding of the role of vibrational coupling in the calcium and strontium phenoxides molecules emerges. 
As summarized in Tables \ref{tab:summary} and \ref{tab:DLIF-vbr}-\ref{tab:DLIF-vib}, except for CaOPh-4-F and SrOPh molecules, all examined molecules show additional vibrational changing decays near the most off-diagonal stretching mode decay ($\nu_k$). 
Specifically, a combination band ($\nu_i\nu_j$) comprising two low-frequency bending modes, which is absent in the harmonic approximation, is activated by anharmonic vibrational coupling. 
This occurs in a predictable manner according to the vibrational frequency spacing and vibrational mode symmetry and can be captured by the VPT calculations.

The strength of this coupling can be estimated from an intensity borrowing model. 
Here, the Fermi resonance Hamiltonian affecting the combination mode $\nu_i\nu_j$ and fundamental mode $\nu_k$ in the ground $\widetilde X$ state can be expressed as~\cite{lefebvre2004spectra}: 
\begin{equation}
    H_{FR, ij, k}^{(\widetilde X)} = \phi_{ij, k}^{(\widetilde X)}\left(\hat{a}_i^{\dagger}\hat{a}_j^{\dagger}\hat{a}_k+\hat{a}_i\hat{a}_j\hat{a}_k^{\dagger}\right),
\end{equation}
where $\phi_{ij, k}^{(\widetilde X)}$ is the coupling strength and $\hat{a}_p$ is the vibrational excitation annihilation operator for mode $p$. 
In the absence of the Fermi resonance (i.e. $\phi_{ij, k}^{(\widetilde X)} = 0$), we assume the probability of decay from the excited state  $|e, \nu'=0\rangle$ to $|\widetilde X, \nu_k''\rangle$ is appreciable, while decay to the combination mode $|\widetilde X, \nu_i''\nu_j''\rangle$ is negligible.
Treating the case of only one combination mode mixing with the stretching mode as a simple two level system, the strength of the vibrational coupling is:
\begin{equation}
    \phi_{ij, k}^{(\widetilde X)}=\frac{\sqrt{\beta_{k/ij}}}{\beta_{k/ij}+1}\sqrt{\Delta \nu_{ij, k}^{(0)}\,^2+4\phi_{ij, k}^{(\widetilde X)}\,^2},
    \label{ratio}
\end{equation} 
where $\beta_{k/ij}$ is the measured intensity ratio of the stretching mode decay to the combination mode decay and $\Delta \nu_{ij, k}^{(0)}$ is the unperturbed energy gap between the modes. 
Using this expression, coupling strengths are extracted and shown in Table~\ref{tab:summary}.
For this comparison, though the unperturbed gap $\Delta \nu_{ij, k}^{(0)}$ could be evaluated from measurement and the above equations, we employ the calculated VPT frequencies for a straightforward comparison of calculated and measured gaps. 

Although Fermi resonance occurring between multiple vibrational modes ($\nu_5$, $\nu_2\nu_3$, $\nu_2\nu_4$) is observed in CaOPh-3-F, evaluating the anharmonic coupling strengths between these modes is challenging.
This is due to the absence of phase factors in the measurement of the off-diagonal matrix elements and state vectors.
Further, the  mixing ratios computed from the intensity borrowing model can produce large uncertainties in these matrix elements as the solution is not unique. 
As a result, the measurement of coupling coefficients of CaOPh-3-F is not available in Table \ref{tab:summary}.  

\begin{table*}
    \centering
    \caption{Summary of Fermi resonance for the most off-diagonal stretching modes in all studied molecules. All frequencies and coupling strengths are given in units of \cm.  }
    \setlength{\tabcolsep}{6pt}
    \begin{tabular}{cccccccccc}
    \hline
    \hline
    \multirow{2}{*}{Species}    & \multicolumn{5}{c}{Theo. (VPT)}       & & \multicolumn{3}{c}{Exp.} \\
    \cline{2-6} \cline{8-10}
                &$\nu_i$ & $\nu_j$  & $\nu_k$  & $\Delta \nu_{ij, k}^{(0)}$ & $\Delta \nu_{ij, k}$ && $\Delta \nu_{ij, k}'$ & $\beta_{k/ij}$ &$\phi_{ij, k}^{(\widetilde X)}$\\
    
    \hline
    CaOPh      & 60.0 (B$_1$, $\nu_2$) &  241.0 (B$_1$, $\nu_3$) &  315.2 (A$_1$, $\nu_4$) & 14.2 & 19.6 &             & 18.0(0.4)           & 2.9(0.8)  &    7.9(0.6)            \\
    CaOPh-3-F   &  57.9 (A$''$, $\nu_2$) &   234.5 (A$''$, $\nu_3$) &  286.0 (A$'$, $\nu_5$) &6.4 & 8.2&              & 6.1(0.6)    & 3.0(1.0)   &     -        \\
               & 57.9 (A$''$, $\nu_2$) &  242.3 (A$''$, $\nu_4$) &  286.0 (A$'$, $\nu_5$) & 14.2 & 16.0 &             & 17.4(0.6) &1.9(0.4)  & -     \\
    CaOPh-3,4,5-F$_3$ &  48.9 (B$_1$, $\nu_1$) &   217.2 (B$_1$, $\nu_4$) &  271.2 (A$_1$, $\nu_6$) & 5.1 & 9.6 &             & 8.2(0.4)            & 1.0(1.0)      &   4.2(2.4)        \\
    SrOPh-3,4,5-F$_3$ &  45.7 (B$_1$, $\nu_1$) & 143.6 (B$_1$, $\nu_3$)&  203.5 (A$_1$, $\nu_4$) & 14.2& 16.2 &            & 18.7(1.0)           &   9.0(4.0)     &     6.0(2.6)      \\
    CaOPh-4-F   & 48.6 (B$_1$, $\nu_1$) &  171.0 (B$_1$, $\nu_3$) &  288.5 (A$_1$, $\nu_4$) &68.9& None&             & \multicolumn{3}{c}{No splitting observed}    \\
    SrOPh      & 52.4 (B$_1$, $\nu_1$) &  54.5 (B$_2$, $\nu_2$)  &  238.6 (A$_1$, $\nu_3$) &131.7& None&           & \multicolumn{3}{c}{No splitting observed}     \\
    \hline
    \hline
    \end{tabular}
    \begin{tablenotes}
    \item Notes: $\nu_i$ and $\nu_j$ are two low-frequency out-of-plane bending modes. The combination band of $\nu_i\nu_j$, FCF-inactive mode under harmonic approximation, is likely to show up due to the intensity borrowing from Fermi resonance coupling with the most-off diagonal stretching mode $\nu_k$ based on the frequency splitting and symmetry.  $\Delta \nu_{ij, k}^{(0)} = |\nu_k-\nu_i-\nu_j|$ is the unperturbed frequency splitting, and $\Delta \nu_{ij, k}=|\nu_k-\nu_i\nu_j|$ are the predicted Fermi resonance splittings ('None' indicates no Fermi resonance for the mode $\nu_k$). The difference of $|\Delta \nu_{ij, k}^{(0)} - \Delta \nu_{ij, k}|$ indicates the frequency shift due to Fermi resonance. All frequencies are calculated at the anharmonic-VPT level of theory. $\Delta \nu_{ij, k}'$ is the measured frequency splitting between the combination band and the stretching mode. $\beta_{k/ij}$ is the averaged measured peak intensity ratio of the stretching mode to the combination band in $\widetilde A \rightarrow \widetilde X$ and $\widetilde B \rightarrow \widetilde X$ transitions. $\phi_{ij, k}^{(\widetilde X)}$ is the estimated Fermi resonance coupling strength between the combination band and the stretching mode in the ground state according to equation (\ref{ratio}). Due to the complexity of coupling between multiple vibrational bands, the coupling strength of CaOPh-3-F could not be estimated from the measurement.  
    \end{tablenotes}
\label{tab:summary}
\end{table*}

The observed anharmonic couplings have substantial implications for the laser cooling of these molecules. 
The presence of additional vibrational decay pathways requires the use of additional repumping lasers to achieve efficient photon scattering~\cite{mccarron2018laser,fitch2021laser,augenbraun2023direct}.
Therefore, it is crucial to design molecules that can minimize or avoid such resonant couplings. 
Several such strategies for mitigating vibrational anharmonic coupling are readily apparent in these molecules. 
First, the spacing of vibrational energy levels can be tailored to maintain sufficient separation of \textit{harmonic states} to avoid detrimental Fermi resonances.
This can be achieved via several approaches, such as substituting groups on the phenyl ring (e.g. CaOPh-4-F) or altering the metal atom hosting the optical cycling center (e.g. SrOPh). 
For example, according to theoretical calculations, it is anticipated that CaOPh-4-Cl, CaOPh-4-OH, SrOPh-3-F and SrOPh-3-OH will not exhibit Fermi resonance coupling between the stretching mode and the bending mode combination band due to their large frequency spacings, as indicated by values exceeding $>60$ \cm, (Table~\ref{tab:predict-molecules}). 
Second, choosing molecules with higher symmetry may protect the stretching mode from mixing with other nearby combination modes, as Fermi resonance only affects modes in the same symmetry.  

As molecular size and complexity increase above the molecules studied here, the increased density of vibrational dark states from the increasingly diverse molecular structure will pose challenges for the effectiveness of the mitigation methods discussed here. 
Selecting suitable ligands with strong electron-withdrawing capability can offer a general suppression of Fermi resonance and higher order couplings. For these molecules, as the molecular orbitals are highly separated from the vibrational degrees of freedom \cite{Dickerson2021FranckCondon,zhu2022caoph}, 
the anharmonic effects induced by these molecular orbitals can be mitigated, therefore the couplings relative to the most off-diagonal modes are suppressed. 

In summary, we have studied Fermi resonance coupling of calcium and strontium phenoxides and their derivatives, employing  high-resolution dispersed laser-induced fluorescence and excitation spectroscopy. 
Fermi resonance phenomena were observed in the ground and excited states for CaOPh, CaOPh-3-F, CaOPh-3,4,5-F$_3$, and SrOPh-3,4,5-F$_3$ molecules. This resonance led to intensity borrowing, particularly in vibrational combination bands consisting of two low-frequency bending modes close in energy to a stretching mode. 
The Fermi resonance effect was absent in CaOPh-4-F and SrOPh due to large frequency differences between the combination band and the stretching mods. While Fermi resonance does not significantly alter vibrational branching ratios, it does require additional repumping lasers for effective optical cycling. 
Several strategies were presented to minimize the impact of Fermi resonance in phenoxide-related molecules, including ligand substitutions and changes in metal atoms. 
These findings help to provide a roadmap for the design and engineering of ever-larger and more intricate molecular systems with enhanced optical cycling properties for advancing quantum information science.

\emph{Acknowledgements --} This work was supported by the AFOSR (grant no. FA9550-20-1-0323), the NSF (grant no. OMA-2016245, PHY-2207985 and DGE-2034835), NSF Center for Chemical Innovation Phase I (grant no. CHE-2221453) and the Gordon and Betty Moore Foundation (DOI: 10.37807/GBMF11566). Computational resources were provided by ACCESS and UCLA IDRE.  C.E.D. thanks Mark Boyer for helpful discussions.

\bibliography{ref}

\newpage

\section*{Supplementary information}

\subsection{Experimental methods}
The experiments were conducted within a cryogenic buffer-gas cell operated at a temperature range of 20-25 K, as depicted in Fig. \ref{fig:setup}. Calcium (or strontium) phenoxide and its derivatives were generated by reacting metal atoms with various organic precursors, including phenol, 3-fluorophenol, 4-fluorophenol, and 3,4,5-trifluorophenol, purchased from Sigma Aldrich. Briefly, an Nd:YAG laser (Minilite) operating at 1064 nm with a pulse energy of approximately 6 mJ and a repetition rate of 10 Hz was employed to ablate calcium or strontium metal pellets, generating excited metal atoms. To prevent production yield drifts, the focused spot of the ablation laser was continuously swept over the target using a moving mirror. These excited metal atoms then reacted with organic ligands introduced into the cryogenic cell through a heated gas line originating from a heated reservoir. Each organic ligand was associated with a separate reservoir. The reaction products were subsequently cooled to their vibrational ground states through collisions with neon buffer gas, with a density of approximately $\approx 10^{15-16}$~\cm. Upon reaching the excitation zone, a tunable pulsed dye laser (LiopStar-E dye laser, operating at 10 Hz) with a linewidth of 0.04 \cm~at 620 nm was utilized to excite molecules to their excited states. Molecules in the excited states underwent spontaneous emission, resulting in the emission of fluorescence. This emitted fluorescence was collected by a lens system and directed into a monochromator (McPherson model 2035) equipped with a 1200 lines/mm grating. Finally, the fluorescence was detected by a photomultiplier tube (PMT).

To investigate the vibrational peak splitting caused by Fermi resonance, we conducted two spectroscopic measurements: dispersed laser-induced fluorescence (DLIF) spectroscopy and excitation spectroscopy. In the DLIF measurement, the laser wavelengths were fixed at the transitions to the vibrational ground level of the electronically excited states, and the fluorescence was dispersed by scanning the grating of the monochromator with an increment of 0.05-0.10 nm. At each grating position, an accumulation of 200-500 shots was taken to ensure reliable signal acquisition. To improve the resolution from the previous measurement \cite{zhu2022caoph,lao2022sroph}, where the spectrometer had a resolution of approximately 0.50 nm, narrower slit widths were used. The entrance slit was set at 0.05 mm, while the exit slit was adjusted to 0.03 mm, achieving a better resolution of 0.20 nm (equivalent to ~ 5.5 \cm). To probe the vibrational decays with low branching ratios, we employed a high laser intensity ($\approx 0.2$ mJ/pulse) in the DLIF measurement. However, this elevated intensity could potentially saturate the 0-0 emission. To accurately calibrate the vibrational branching ratios, a lower-intensity laser ($\approx 0.02$ mJ/pulse) was used to measure the relative ratio of the 0-0 peak and the most off-diagonal stretching mode peak in the ground state $\widetilde X$. This ratio was used to scale down the 0-0 peak in the DLIF measurement under high laser intensity.

The excitation spectroscopy aimed to detect the vibrational splitting in the excited states $\widetilde A$ and $\widetilde B$. During this process, the pulsed dye laser wavelengths were scanned at an increment of 0.02-0.10 nm for the $\widetilde A (v_n')/\widetilde B (v_n') \leftarrow \widetilde X (v'' = 0)$, while the grating position remained fixed at the corresponding 0-0 transition. This allowed us to explore the off-diagonal excitation while simultaneously monitoring the diagonal emission.

\renewcommand*{\thefigure}{S\arabic{figure}}
\setcounter{figure}{0}
\begin{figure}
    \centering
    \includegraphics[scale=0.4]{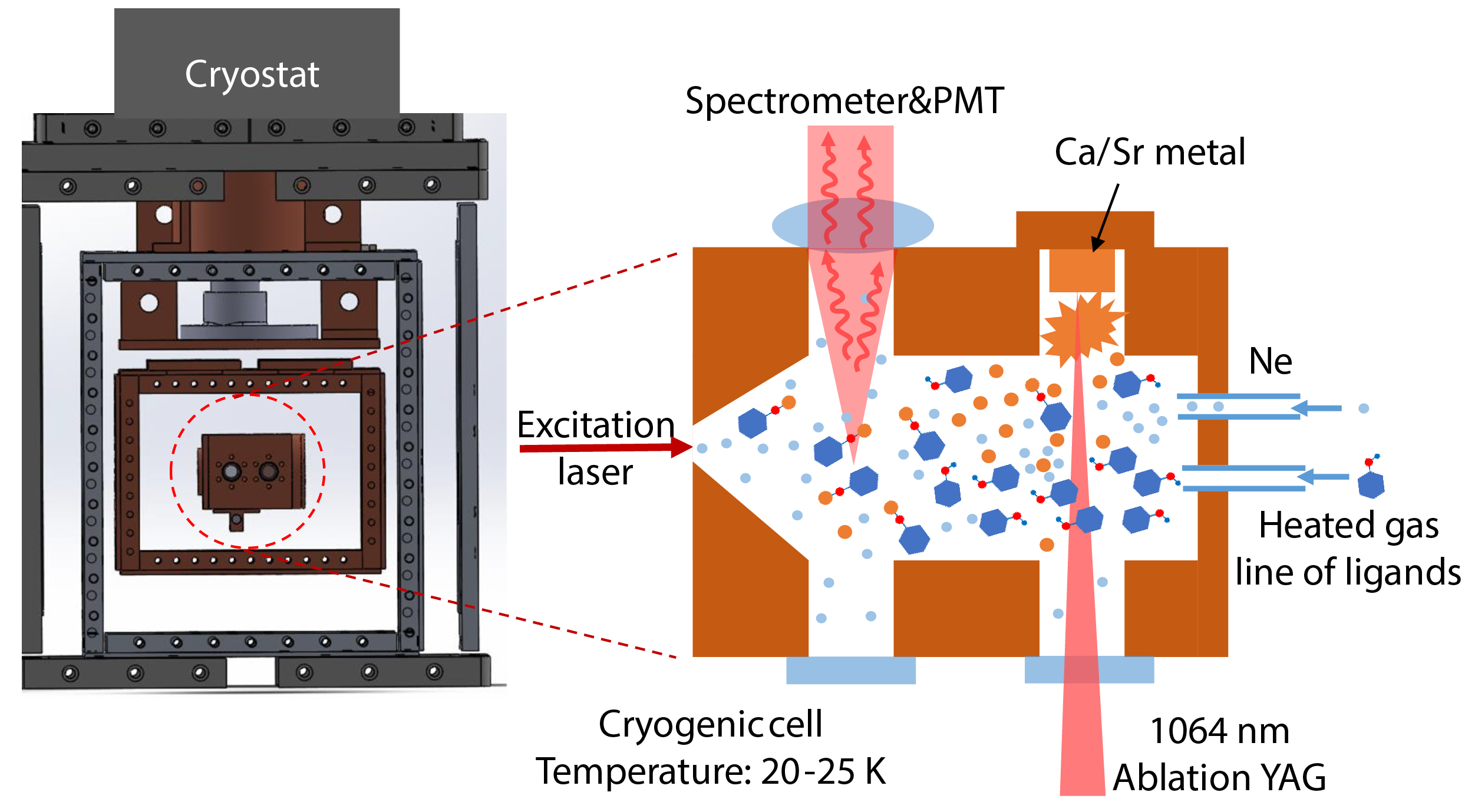}
    \caption{Schematic illustration of the experiment setup. } 
    \label{fig:setup}
\end{figure}

\subsection{Theoretical methods}
All calculations were performed at the PBE0-D3/def2-TZVPPD level of theory~\cite{perdew1996rationale, weigend2005balanced, grimme2010consistent,rappoport2010property} with a superfine grid in Gaussian16~\cite{frisch2016gaussian}. An isosurface of 0.03 was used to generate all molecular orbitals with the Multiwfn program~\cite{lu2012multiwfn}. 
 Optimized geometries, vertical excitation energies and frequencies were calculated with density functional theory (DFT)/time-dependent DFT (TD-DFT) methods.  Harmonic Franck Condon factors (FCFs) were obtained using the harmonic approximation with Duschinsky rotations up to three quanta in ezFCF~\cite{ezFCF2022}.  Anharmonic frequencies and anharmonic-corrected FCFs were calculated with vibrational perturbation theory (VPT)~\cite{VPT1nielsen1951vibration,VPT2barone2005anharmonic} using $PyVPT_{n}$~\cite{pyVibPtboyer2021,boyer2022}.    
 
As seen in this work, anaharmonic coupling that leads to intensity-borrowing is missed by the harmonic approximation.  To predict Fermi resonances and anharmonic-corrected FCFs, we use the numerical, matrix-form VPT as implemented in $PyVPT_{n}$ with the full $3N-6$ mode basis. 
Gaussian16 was used to obtain the quartic expansion of the normal mode potentials by evaluating the first and second derivatives of the Hessian at the PBE0-D3/def2-TZVPPD level of theory.   A wavefunction threshold of 0.3 and energy threshold of 500~\cm was used for identifying degenerate subspaces, based on past investigations of these thresholds~\cite{wfcthresholds2022}.  

Since VPT coupling matrices are sensitive to small changes in the diagonal energies, and diagonal energies are based on the quality of the original Hamiltonian initial inputs of harmonic frequencies and quartic expansions, some frequencies were shifted up to 7~\cm based on experimental evidence, as done in past work~\cite{depertshifts12022,anharmFCFs2023}.  For CaOPh, the $2\nu_4$ diagonal perturbed anharmonic frequency was shifted $2\nu_4+6.8$~\cm.  This shift is incorporated in the resulting coupling matrices in Table \ref{tab:coupling}, whose diagonalized matrices gave the corrected frequencies and coefficients used for anharmonic FCFs, as explained in Section III.  For CaOPh-3-F, the original harmonic $\nu_6$ was shifted by $\nu_6-2$~\cm, but no deperturbed anharmonic frequencies were shifted. For all other molecules, no shifts were made.

\subsection{Anharmonic Franck-Condon Factors}\label{sec:anharmonic}

We adopt the same method used in Ref. \cite{anharmFCFs2023} to include anharmonic corrections based on harmonic FCFs.

Anharmonic vibrational eigenstates are given by:
\begin{equation}
\ket{\chi''} = \sum_{j}c''_{j}\ket{\Phi''_{j}}
\end{equation}
where $c_{j}''$ are the eigenstates from the diagonalized VPT coupling matrices (see Table \ref{tab:coupling}) and $\ket{\Phi''_{j}}$ represents the zeroth-order state basis used in VPT.

Anharmonic FCFs are calculated as a transition form some initial state, j, to final state, k, as:
\begin{equation}
    |\bra{\chi'}\ket{\chi''}|^{2} = |\sum_{j,k}c'_{k}c''_{j}\bra{\Phi'_{k}}\ket{\Phi''_{j}}|^{2}
\end{equation}

We approximate the zeroth-order wavefunctions used in VPT are approximately the harmonic normal mode wavefunctions, $\ket{\Phi}\approx \ket{\Phi_{h}}$.  This is expected to be a good approximation because $\ket{\Phi}$ are from the deperturbed VPT calculations, which makes other state contributions small compared to the leading term in the expansion. 
The matrix is then diagonalized to get obtain full state mixing contributions which are incorporated via the mixing coefficients, $|c|^{2}$. This gives the revised equation, below:
\begin{equation}
    |\bra{\chi'}\ket{\chi''}|^{2} = |\sum_{j,k}c'_{k}c''_{j}\bra{\Phi'_{h,k}}\ket{\Phi''_{h,j}}|^{2}
    \label{eq:ho}
\end{equation}

Since our excited-state molecule is at its vibrational ground state, we can approximate the initial state as one eigenstate, $\ket{\chi''}\approx \ket{\Phi''_{h,j}}$, so that the FC factor arising from the zeroth-order excited state only involves a single overlap integral between two harmonic states, which is computed analytically in ezFCF:
\begin{equation}
|\sum_{j}c''_{j}\bra{\Phi'_{k}}\ket{\Phi''_{j}}|^{2} = |\bra{\Phi'_{k}}\ket{\Phi''_{j}}|^{2}
\end{equation}

Using these approximations, anharmonic-corrected FCFs, which we report in Table \ref{tab:DLIF-vbr}, are built using harmonic wavefunctions ($\ket{\phi_{h}}$) as a basis and mixing coefficients ($c'_{k}$) obtained from VPT, using the final equation below:
\begin{equation}
|\bra{\chi'}\ket{\chi''}|^{2} \approx |c'_{k}|^{2}|\bra{\Phi'_{h,k}}\ket{\Phi''_{h,j}}|^{2}
\end{equation}

\subsection{Error analysis of vibrational branching ratios}
All peaks observed in the DLIF spectra were fitted with the Gaussian function and the peak areas were extracted to estimate the respective vibrational branching ratios (VBRs) \cite{lao2022sroph}. The corresponding statistical uncertainty is calculated with the following formula

\begin{equation}
\begin{split}
S_i = I_i/\sum_{j=0}^pI_j
\end{split}
\end{equation}

\begin{equation}
\begin{split}
    \Delta S_i = &\sqrt{\sum_{j=0}^p\left(\frac{\partial S_i}{\partial I_j}\right)^2\Delta I_j^2}\\
    =&\frac{1}{S}\sqrt{(1-2S_i)\Delta I_i^2+S_i^2\sum_{j=0}^p\Delta I_j^2},
\end{split}
\end{equation}
where $S_i$ and $\Delta S_i$ are the VBR and VBR uncertainty of each observed vibrational peak $i$. $p$ is the number of all observed peaks. $I_i$ and $\Delta I_i$ are the intensity (or area) and intensity uncertainty of peak $i$ from the Gaussian fitting. $S=\sum_{j=0}^pI_j$ is the total intensity of all observed vibrational decays. The VBR results are summarized in Table \ref{tab:DLIF-vbr}.

The systematic error sources are mainly from the unobserved vibrational decays, signal drifting in the measurement, spectrometer response of the fluorescence detection and the diagonal excitations, as discussed in previous work\cite{lao2022sroph}. For simplicity, the updated errors are summarized in Table \ref{Syserr}.

The PMT may be saturated if the fluorescence signal is too strong, especially during the detection of diagonal decay signals. This saturation can lead to a decrease in the measured signal strength and lower diagonal VBRs. To address this issue, the DLIF scan is repeated at different excitation powers. The off-diagonal decays are measured from scans with high excitation power, while the diagonal decay signal strength is restored by scaling the scans with low excitation power. The scaling factors are determined by the average ratios of the off-diagonal peak strengths measured in both scans. This scaling method is applied when significant discrepancies in VBRs are observed between scans. The scaling process can introduce larger uncertainties in the diagonal VBRs compared to the off-diagonal VBRs, as shown in the table, due to larger relative uncertainties in the intensities of the off-diagonal decays obtained under weak excitation. 

\renewcommand*{\thetable}{S\arabic{table}}
\setcounter{table}{0}

\begin{table*}
\caption{Ground state vibrational perturbation theory coupling matrices involving the deperturbed frequencies in \cm. 
 $\alpha$ denotes a deperturbed frequency shift.}
\centering
\setlength{\tabcolsep}{7pt}
\renewcommand{\arraystretch}{1.2}
\label{tab:coupling}
\begin{tabular}{cccccccccc}
\hline
\hline
\multicolumn{9}{c}{CaOPh}  \\ 
\hline
& $\nu_4$  & $\nu_2\nu_3$ &  &  & $\nu_9$ & $2\nu_4$ & $\nu_2\nu_3\nu_4$ \\
$\nu_4$    &    $311.052$    &  -7.999 &                &  $\nu_9$&    $624.025$   & -7.162  &  -1.06           \\
$\nu_2\nu_3$  &   -7.999   & $299.719$  &              &  2$\nu_4$  &  -7.162  &  $628.079^{\alpha}$   &   -11.313           \\
&      &   &  &   $\nu_2\nu_3\nu_4$ & -1.06 & -11.313 & 653.769 &  \\

\hline
\multicolumn{9}{c}{CaOPh-3-F}  \\ 
\hline
& $\nu_5$  & $\nu_2\nu_3$ & $\nu_2\nu_4$ &  & $\nu_6$ & $\nu_1\nu_5$ & $\nu_1\nu_2\nu_4$ \\
$\nu_5$ & 291.107 & -2.821 & -6.298 
& $\nu_6$ & 357.272 & -3.441 & -0.509\\
$\nu_2\nu_3$ & -2.821 & 292.892 &  -1.356 
& $\nu_1\nu_5$ & -3.441 & 342.542 & -6.298\\
$\nu_2\nu_4$ & -6.298 & -1.356 & 298.489
& $\nu_1\nu_2\nu_4$ & -0.509 & -6.298 & 351.191\\
\hline
\multicolumn{3}{c}{CaOPh-3,4,5-F$_3$} &  & & & \multicolumn{3}{c}{SrOPh-3,4,5-F$_3$} \\ 
\cline{1-3} \cline{7-9}
& $\nu_6$ & $\nu_1\nu_4$ &  & & & & $\nu_4$ & $\nu_1\nu_3$ \\
$\nu_6$ & 267.266 & 4.704 &  & & &  $\nu_4$ & $201.611$ & 5.207\\
$\nu_1\nu_4$ & 4.704	& 265.535 &  & & & $\nu_1\nu_3$ & 5.207 & $189.229$\\
\hline
\hline
\end{tabular}
\end{table*}

\begin{table*}
    \centering
    \caption{Summary of systematic errors in estimating the vibrational branching ratios.}
    \setlength{\tabcolsep}{7pt}
    \renewcommand{\arraystretch}{1.2}
    \begin{tabular}{c c c c c}
    \hline
    \hline
        Instrument & Signal  & Unobserved & Diagonal & Total\\
         response & drifting & peaks & excitations & error\\
        \hline
         1\% & 1\% & 3\% & 0.5\% & 4\%\\
        \hline
        \hline
    \end{tabular}
    \label{Syserr}
\end{table*}

\begin{table*}
\caption{Experimental vibrational branching ratios and theoretical FCFs of $\widetilde A \rightarrow \widetilde X$ and $\widetilde B \rightarrow \widetilde X$ transitions in DLIF spectra (Figs. \ref{fig:CaOPh_DLIF}-\ref{fig:SrOPh-DLIF}, \ref{fig:345F-DLIF} and \ref{fig:sroph345f-dlif}) of all molecules studied in this work. The values in parentheses are standard errors in the Gaussian fit to extract the peak areas \cite{zhu2022caoph,lao2022sroph}, not including the systematic uncertainties. The theoretical FCFs are obtained under the harmonic approximation (Harm.) and anharmonic-corrected method based on vibrational perturbation theory (Anharm. VPT).  The vibrational modes are ordered with increasing frequency. The label * indicates the vibrational modes with Fermi resonance coupling.}
\label{tab:DLIF-vbr}
\centering
\setlength{\tabcolsep}{4pt}
\renewcommand{\arraystretch}{1}
\resizebox{!}{0.32\paperheight}{
\begin{tabular}{ccccccc}
\hline
\hline
\multirow{2}{*}{Modes} & \multicolumn{6}{c}{CaOPh}                                    \\ \cline{2-7}
                       & Exp. ($\widetilde A$ )    & Harm. ($\widetilde A$ )     & Anharm. VPT ($\widetilde A$ )   & Exp. ($\widetilde B$ )    & Harm. ($\widetilde B$ )   & Anharm. VPT ($\widetilde B$ )     \\ \hline
0                      & 0.930(30)  & 0.9575   &   --    & 0.909(22)  & 0.9736 & --        \\
$\nu_2$                   & 0.009(4) & \textless10$^{-4}$ & -- & 0.009(3) &\textless10$^{-4}$ & -- \\
$\nu_3$                       & --           &       \textless10$^{-4}$      &  --  & 0.002(2) & \textless10$^{-4}$ & -- \\
$\nu_2\nu_3*$                    & 0.013(6) & \textless10$^{-4}$ & 0.0071 & 0.021(5) & \textless10$^{-4}$ & 0.0044\\
$\nu_4*$                      & 0.043(18)  & 0.0329          & 0.0264 & 0.049(12)  & 0.0196   & 0.0158       \\
2$\nu_4*$                      & 0.002(2) &0.0007 & 0.0016 & 0.004(3) & 0.0003 & 0.0014\\
$\nu_9*$                       & 0.003(2) & 0.0030       &  0.0019 & 0.006(3) & 0.0031    & 0.0017      \\
\hline
\multirow{2}{*}{Modes} & \multicolumn{6}{c}{CaOPh-4-F}                                   \\ \cline{2-7}
                       & Exp. ($\widetilde A$ )    & Harm. ($\widetilde A$ )     & Anharm. VPT ($\widetilde A$ )   & Exp. ($\widetilde B$ )    & Harm. ($\widetilde B$ )   & Anharm. VPT ($\widetilde B$ )     \\ \hline
0                      & 0.902(7)   & 0.9614         & --       & 0.928(4) &  0.9773       & --        \\
$\nu_2$                       & 0.008(5)   & \textless10$^{-4}$     & --           & 0.004(2) &  \textless10$^{-4}$   & --            \\
$\nu_3$                       &--     &\textless10$^{-4}$  & --               & 0.005(2) &  \textless10$^{-4}$    & --           \\
$\nu_4$                      & 0.088(4)   & 0.0297        & --        & 0.060(3) & 0.0165       & --         \\
$\nu_2\nu_4$                     & 0.002(4)   & \textless10$^{-4}$    & --            &0.003(9)            &\textless10$^{-4}$          & --       \\
\hline
\multirow{2}{*}{Modes} & \multicolumn{6}{c}{CaOPh-3-F}                                   \\ \cline{2-7}
                       & Exp. ($\widetilde A$ )    & Harm. ($\widetilde A$ )     & Anharm. VPT ($\widetilde A$ )   & Exp. ($\widetilde B$ )    & Harm. ($\widetilde B$ )   & Anharm. VPT ($\widetilde B$ )     \\ \hline
0                      & 0.917(13)  & 0.9645    &   --  & 0.901(11)  & 0.9806    & --      \\
$\nu_1$                      & 0.016(7)   & 0.0009 &     --      & 0.009(5)   & \textless10$^{-4}$ & -- \\
$\nu_5*$                 & 0.029(7)   & 0.0234     & 0.0150     & 0.048(7)   & 0.0129   & 0.0083       \\
$\nu_2\nu_3*$                & 0.013(6)   & \textless{}10$^{-4}$ & 0.0042 & 0.013(5)   & \textless{}10$^{-4}$ & 0.0023\\
$\nu_2\nu_4*$                    & 0.015(5)   & \textless{}10$^{-4}$ & 0.0049 & 0.025(4)   & \textless{}10$^{-4}$ & 0.0028\\
$\nu_1\nu_5*$ &-- & \textless{}10$^{-4}$ & 0.0003 & --& \textless{}10$^{-4}$ & 0.0002\\
$\nu_6*$                    & 0.010(6)   & 0.0033             & 0.0028   & 0.004(4)& 0.0014 & 0.0012       \\
\hline
\multirow{2}{*}{Modes} & \multicolumn{6}{c}{CaOPh-3,4,5-F$_3$}                             \\ \cline{2-7}
                       & Exp. ($\widetilde A$ )    & Harm. ($\widetilde A$ )     & Anharm. VPT ($\widetilde A$ )   & Exp. ($\widetilde B$ )    & Harm. ($\widetilde B$ )   & Anharm. VPT ($\widetilde B$ )     \\ \hline
0                      & 0.918(9) & 0.9732      &  --  & 0.958(39)  & 0.9875   & --       \\
$\nu_1$                      & 0.005(7) & \textless{}10$^{-4}$ &   --   &   --   & \textless{}10$^{-4}$ & --\\
$\nu_1\nu_4*$                    & 0.030(5) & \textless{}10$^{-4}$ &  0.0069 & 0.018(28)  & \textless{}10$^{-4}$ & 0.0031 \\
$\nu_6*$                    & 0.030(4) & 0.0167     &   0.0099   & 0.017(27)  & 0.0076 & 0.0044         \\
$\nu_8$                     & 0.017(3) & 0.0033    &   --    & 0.007(8)   & 0.0008   & --       \\
\hline
 
\multirow{2}{*}{Modes}   & \multicolumn{6}{c}{SrOPh} \\ \cline{2-7}
                       & Exp. ($\widetilde A$ )    & Harm. ($\widetilde A$ )     & Anharm. VPT ($\widetilde A$ )   & Exp. ($\widetilde B$ )    & Harm. ($\widetilde B$ )   & Anharm. VPT ($\widetilde B$ )     \\ 
\hline
0           &0.888(11)          &0.9325     &   --   & 0.872(13)         &   0.9497 & --\\
$\nu_2$      & 0.015(9)         &\textless{}10$^{-4}$       &  --  & 0.016(11)         & \textless{}10$^{-4}$  & --\\
$\nu_3$     &0.097(7)   &0.0564   &  --   &  0.112(9)            &     0.0416   & --    \\
\hline
\multirow{2}{*}{Modes} & \multicolumn{6}{c}{SrOPh-3,4,5-F$_3$}                             \\ \cline{2-7}
                       & Exp. ($\widetilde A$ )    & Harm. ($\widetilde A$ )     & Anharm. VPT ($\widetilde A$ )   & Exp. ($\widetilde B$ )    & Harm. ($\widetilde B$ )   & Anharm. VPT ($\widetilde B$ )     \\ \hline
0           &0.929(14) &0.9477      &  -- &0.924(7) &0.9621                     & --   \\
$\nu_2$     &0.010(10)  &\textless{}10$^{-4}$     &  --    &0.007(3) & \textless{}10$^{-4} $  & --                      \\
$\nu_3$     & --  & \textless{}10$^{-4}$     &  --    &0.017(5) & \textless{}10$^{-4}$   & --                     \\
$\nu_1\nu_3*$     &0.011(9)  &\textless{}10$^{-4}$    & 0.0051       &0.004(2) & \textless{}10$^{-4}$ & 0.0034                        \\
$\nu_4*$     &0.050(6)  &0.0422   & 0.0382       &0.049(3) &  0.0285 & 0.0256                      \\ 
\hline
\hline
\end{tabular}}
\end{table*}

\begin{table*}
\caption{Resolved vibrational modes for the ground states in the DLIF spectra (Figs. \ref{fig:CaOPh_DLIF}-\ref{fig:SrOPh-DLIF}, \ref{fig:345F-DLIF} and \ref{fig:sroph345f-dlif}). The values in the parentheses are statistical errors when fitting resolved peaks with Gaussian functions. The theoretical harmonic frequencies (Harm. freq.) and anharmonic frequencies (VPT freq.) are given for comparison. The harmonic frequencies of combination or overtone bands are simply the sum of frequencies from individual vibrational modes. The label * indicates the vibrational modes with observed Fermi resonance coupling. All values are in units of \cm.  }
\centering
\setlength{\tabcolsep}{6.5pt}
\renewcommand{\arraystretch}{1.2}
\label{tab:DLIF-vib}
\resizebox{!}{.25\paperwidth}{%
\begin{tabular}{cccccccccc}
\hline
\hline
\multicolumn{4}{c}{CaOPh}                          &  & \multicolumn{4}{c}{SrOPh}                          \\ \cline{1-4} \cline{6-9} 
Vib. modes & Exp. freq        & Harm. freq. & VPT freq. &   & Vib. modes & Exp. freq          & Harm. freq. & VPT freq.  \\
 \cline{1-4} \cline{6-9} 
$\nu_2$         & 59.8 (0.7)       & 61  & 60.0               &  & $\nu_2$         & 54.4 (1.2)       & 56.4 & 54.5             \\
$\nu_3$         & 240.9 (3.1)      & 246.5  & 241.0            &  & $\nu_3$         & 235.5 (0.1)      & 240.5 &  238.6            \\
$\nu_2\nu_3$*       & 294.6 (0.4)      & 307.5    & 295.6         &  & &&&\\              
$\nu_4$*         & 312.6 (0.1)      & 313.6 & 315.2             &  & &&&\\
2$\nu_4$*        & 621.2 (1.3)      & 627.3             &  613.1&            &                  &                    \\
$\nu_9$*         & 630.3 (1.0)      & 630.5             &  628.7&            &                  &                    \\ 
\hline

\multicolumn{4}{c}{CaOPh-3-F}                      &  & \multicolumn{4}{c}{CaOPh-4-F}                      \\ \cline{1-4} \cline{6-9} 
Vib. modes & Exp. freq        & Harm. freq. & VPT freq. &   & Vib. modes & Exp. freq          & Harm. freq. & VPT freq.  \\
 \cline{1-4} \cline{6-9} 
$\nu_1$         & 54.4 (1.0)       & 58.8 & 53.8               &  & $\nu_2$         & 52.9 (1.2)       & 57.4  & 55.3             \\
$\nu_5$*         & 284.4 (0.4)      &  296.7 & 286.1             &  & $\nu_3$         & 169.6 (0.7)      & 175.9    & 171.0          \\
$\nu_2\nu_3$*       & 290.5 (0.4)      & 296.8 & 294.3             &  & $\nu_4$         & 285.9 (0.1)      & 291.1     & 288.5         \\
$\nu_2\nu_4$*       & 301.8 (0.5)      & 307.9 & 302.1             &  & $\nu_2\nu_4$       & 346.3 (6.2)      &   348.5 & 343.8                \\
$\nu_1\nu_5$* & -- & 351.6 & 338.6\\

$\nu_6$*         & 355.3 (1.3)      & 359.2   & 358.4           &  &            &                  &                    \\ 
\hline
\multicolumn{4}{c}{CaOPh-3,4,5-F$_3$}                    &  & \multicolumn{4}{c}{SrOPh-3,4,5-F$_3$}                    \\ \cline{1-4} \cline{6-9} 
Vib. modes & Exp. freq        & Harm. freq. & VPT freq. &   & Vib. modes & Exp. freq          & Harm. freq. & VPT freq.  \\
 \cline{1-4} \cline{6-9} 
$\nu_1$         & 50.2 (0.8)       & 50.5               &  48.9 & & $\nu_2$         & 43.9 (1.1)       & 45.4 & 45.9               \\
$\nu_1\nu_4$*       & 261.6 (0.3)      & 272.0  & 261.6             &  & $\nu_3$         & 138.7 (0.9)      & 146.2    & 143.6                \\
$\nu_6$*         & 269.8 (0.2)      & 271.6 & 271.2             &  & $\nu_1\nu_3$*       & 183.1 (1.0)      &   190.5 & 187.3               \\
$\nu_8$         & 309.4 (0.3)      & 314.7    & 313.4        &  & $\nu_4$*         & 201.8 (0.1)      & 204 & 203.5      \\
\hline
\hline
\end{tabular}}
\end{table*}

\begin{table*}
    \centering
    \setlength{\tabcolsep}{5.5pt}
    \renewcommand{\arraystretch}{1.2}
    \caption{The frequencies and assignments of all observed vibrational peaks in the excitation spectra of calcium phenoxides (Figs. \ref{fig:CaOPh_DLIF}, \ref{fig:4F-excitaiton}, \ref{fig:3F-excitation} and \ref{fig:345F-excitation}). The theoretical frequencies are ground-state VPT calcluations.  Vibrational modes involved with the Fermi resonance coupling are labeled with *. The uncertainties of observed frequency shifts are within 5 \cm. All values are in units of \cm. }
    \label{tab:vib-in-AB-Ca}
    \begin{tabular}{ccccccccc}
    \hline
    \hline                                                            
    \multicolumn{4}{c}{CaOPh $\widetilde A \leftarrow \widetilde X$ }  &  & \multicolumn{4}{c}{CaOPh $\widetilde B \leftarrow \widetilde X$ }                         \\ \cline{1-4} \cline{6-9} 
    Observed peak       & Freq. above & Assigned  & VPT   &  & Observed peak       & Freq. above & Assigned & VPT  \\ 
    wavelength & $\widetilde A$  (v=0)     & modes     & freq.  &  & wavelength & $\widetilde B$  (v=0)     & modes    & freq.  \\ \cline{1-4} \cline{6-9} 
    605.60     & 300.1          & $\nu_2\nu_3$*         & 295.6     &   & 600.48     & 300.1          & $\nu_2\nu_3$*             & 295.6     \\
    605.00     & 316.5          & $\nu_4$*           & 315.2     &  & 599.88     & 316.8          & $\nu_4$*               & 315.2     \\
    594.98     & 594.8          & $\nu_2\nu_3\nu_4$*       &    657.4        &  & 590.08     & 593.6          & $\nu_2\nu_3\nu_4$*           &    657.4   \\
    594.28     & 614.6          & 2$\nu_4$*          & 613.1     &  & 589.38     & 613.8          &2$\nu_4$*              & 613.1     \\
    593.88     & 626.0          & $\nu_9$*           & 628.7     &  & 588.98     & 625.3          & $\nu_9$*               & 628.7     \\
    \hline                                                            
    \multicolumn{4}{c}{CaOPh-4-F $\widetilde A \leftarrow \widetilde X$ } &  & \multicolumn{4}{c}{CaOPh-4-F $\widetilde B \leftarrow \widetilde X$ }  \\ \cline{1-4} \cline{6-9} 
    Observed peak       & Freq. above & Assigned  & VPT   &  & Observed peak       & Freq. above & Assigned & VPT  \\ 
    wavelength & $\widetilde A$  (v=0)     & modes     & freq.  &  & wavelength & $\widetilde B$  (v=0)     & modes    & freq.  \\ \cline{1-4} \cline{6-9} 
    604.80     & 286.4          & $\nu_4$           & 288.5     &  & 603.15     & 172.7          & $\nu_3$               & 171.0      \\
    602.65     & 345.4          & $\nu_1\nu_4$         &      337.1     &  & 599.04     & 286.5          & $\nu_4$               & 288.5     \\
    602.50     & 349.5          & $\nu_2\nu_4$         &   343.8         &  & 596.89     & 346.6          & $\nu_2\nu_4$             &     343.8     \\
    602.05     & 361.9          & $\nu_5$            & 364.7      &  & 595.99     & 371.9          & $\nu_5$               & 364.7         \\
    599.89     & 421.7          & $\nu_2\nu_5$         &     420.6       &  & 594.25     & 421.0          & $\nu_2\nu_5$             &    420.6    \\
    598.44     & 462.1          & $\nu_3\nu_4$         &    457.9        &  & 592.80     & 462.2          & $\nu_3\nu_4$             &     457.9     \\
    \hline                                                            
    \multicolumn{4}{c}{CaOPh-3-F $\widetilde A \leftarrow \widetilde X$ }   &  & \multicolumn{4}{c}{CaOPh-3-F $\widetilde B \leftarrow \widetilde X$ }                    \\ \cline{1-4} \cline{6-9} 
    Observed peak       & Freq. above & Assigned  & VPT   &  & Observed peak       & Freq. above & Assigned & VPT  \\ 
    wavelength & $\widetilde A$  (v=0)     & modes     & freq.  &  & wavelength &$\widetilde B$  (v=0)     & modes    & freq.  \\ \cline{1-4} \cline{6-9} 
    602.32     & 289.5          & $\nu_5$*           & 286.1     &  & 597.53     & 288.2          & $\nu_5$*               & 286.1     \\
    602.15     & 294.2          & $\nu_2\nu_3$*         & 294.3     &  & 597.37     & 292.7          & $\nu_2\nu_3$*            & 294.3     \\
    601.65     & 308.0          & $\nu_2\nu_4$*         & 302.1     &  & 596.87     & 306.7          & $\nu_2\nu_4$*              & 302.1     \\
    600.37     & 343.4          & $\nu_1\nu_5$*         &     338.6       &  & 595.49     & 345.5          & $\nu_1\nu_5$*             &    338.6        \\
    600.17     & 349.0          & $\nu_1\nu_2\nu_3$*       &      347.7      &  & 595.33     & 350.0          & $\nu_1\nu_2\nu_3$*           &     347.7       \\
    599.93     & 355.7          & $\nu_6$*/$\nu_1\nu_2\nu_4$*    & 358.4/354.2    &  & 595.09     & 356.8          & $\nu_6$/$\nu_1\nu_2\nu_4$*        & 358.4/354.2     \\
    \hline                                                            
    \multicolumn{4}{c}{CaOPh-3,4,5-F$_3$ $\widetilde A \leftarrow \widetilde X$ }  &  & \multicolumn{4}{c}{CaOPh-3,4,5-F$_3$ $\widetilde A \leftarrow \widetilde X$ } \\ \cline{1-4} \cline{6-9} 
    Observed peak       & Freq. above & Assigned  & VPT   &  & Observed peak       & Freq. above & Assigned & VPT  \\ 
    wavelength & $\widetilde A$  (v=0)     & modes     & freq.  &  & wavelength & $\widetilde B$  (v=0)     & modes    & freq.  \\ \cline{1-4} \cline{6-9} 
    599.49     & 253.7          & $\nu_5$         &  254.8      &  & 601.69     & 49.8           & $\nu_1$               & 48.9     \\
    599.09     & 264.9          & $\nu_1\nu_4$*       &  261.1     &  & 594.49     & 251.1          & $\nu_5$               & 254.8      \\
    598.79     & 273.2          & $\nu_6$*         &  271.2     &  & 594.09     & 262.5          & $\nu_1\nu_4$*             & 261.1     \\
     597.49     & 309.6          & $\nu_8$        &  313.4      &  & 593.74     & 272.4          & $\nu_6$*               & 271.2     \\
     595.79     & 357.3          & $\nu_2\nu_8$*      &  366.0     &  & 593.19     & 288.0          & 2$\nu_3$              &    285.4       \\
     595.59     & 363.0          & $\nu_9$*        &  357.4     &  & 592.44     & 309.3          & $\nu_8$               & 313.4       \\
                &                 &            &             &  & 590.84     & 355.0          & $\nu_2\nu_8$*             & 366.0    \\
                &                 &            &             &  & 590.59     & 362.2          & $\nu_9$*               & 357.4    \\   
    \hline
    \hline    
    \end{tabular}
\end{table*}

\begin{table*}
\centering
\setlength{\tabcolsep}{6pt}
\renewcommand{\arraystretch}{1.2}
\caption{The frequencies and assignments of all observed vibrational peaks in the excitation spectra of strontium phenoxides (Figs. \ref{fig:sroph-excitation} and \ref{fig:sroph345f-excitation}). The theoretical frequencies are ground-state VPT calcluations.  The fundamental vibrational modes involved with the Fermi resonance coupling (labeled with *) are given. The uncertainties of observed frequency shifts are within 5 \cm. The  All values are in units of \cm.  }
\label{tab:vib-in-AB-Sr}
\begin{tabular}{ccccccccc}
\hline
\hline                                                            
\multicolumn{4}{c}{SrOPh $\widetilde A \leftarrow \widetilde X$ }                                 &  & \multicolumn{4}{c}{SrOPh $\widetilde B \leftarrow \widetilde X$ }                         \\ \cline{1-4} \cline{6-9} 
    Observed peak       & Freq. above & Assigned  & VPT   &  & Observed peak       & Freq. above & Assigned & VPT  \\ 
    wavelength & $\widetilde A$ (v=0)     & modes     & freq.  &  & wavelength & $\widetilde B$ (v=0)     & modes    & freq.  \\ \cline{1-4} \cline{6-9} 
667.06     & 58.2           & $\nu_2$          & 54.5       &  & 653.74     & 58.3           & $\nu_2$          & 54.5       \\
659.05     & 240.4          & $\nu_3$          & 238.6      &  & 645.98     & 242.0          & $\nu_3$          & 238.6      \\
656.55     & 298.2          & $\nu_2\nu_3$        &     294.1       &  & 643.73     & 296.1          & $\nu_2\nu_3$        &    294.1   \\
\hline                                                            
\multicolumn{4}{c}{SrOPh-3,4,5-F$_3$ $\widetilde A \leftarrow \widetilde X$ }                                 &  & \multicolumn{4}{c}{SrOPh-345F $\widetilde B \leftarrow \widetilde X$ }                    \\ \cline{1-4} \cline{6-9} 
    Observed peak       & Freq. above & Assigned  & VPT   &  & Observed peak       & Freq. above & Assigned & VPT  \\ 
    wavelength & $\widetilde A$ (v=0)     & modes     & freq.  &  & wavelength & $\widetilde B$ (v=0)     & modes    & freq.  \\ \cline{1-4} \cline{6-9} 
655.44     & 47.8           & $\nu_2$         & 45.9       &  & 642.15     & 44.0           & $\nu_2$            & 45.9       \\
649.33     & 191.4          & $\nu_1\nu_3$*        & 187.3      &  & 636.31     & 186.9     & $\nu_1\nu_3$*       & 187.3      \\
648.73     & 205.6          & $\nu_4$*          & 203.5        &  & 635.61     & 204.2      & $\nu_4$*            & 203.5       \\
\hline
\hline    
\end{tabular}
\end{table*}

\begin{table*}
    \caption{The predicted molecules without Fermi resonance coupling for the most-off diagonal stretching mode $\nu_k$. All units are in \cm}
    \centering
    \setlength{\tabcolsep}{10pt}
    \renewcommand{\arraystretch}{1.2}
    \begin{tabular}{c|cccc}
    \hline
    \hline
    Species    & $\nu_i$ & $\nu_j$  & $\nu_k$  & $\Delta \nu_{ij, k}^{(0)}$ \\
    \hline
    CaOPh-4-Cl & 43.7 (B$_1$, $\nu_1$) & 142.6 (B$_1$, $\nu_3$) & 263.0 (A$_1$, $\nu_4$) & 76.7       \\
    CaOPh-4-OH & 54.6 (A$''$, $\nu_1$) & 172.1 (A$''$, $\nu_3$) & 291.9 (A$'$, $\nu_4$) & 65.2      \\
    SrOPh-3-F  & 48.4 (A$'$, $\nu_1$) & 53.7 (A$''$, $\nu_2$)  & 224.8 (A$'$, $\nu_3$) & 122.7      \\
    SrOPh-3-OH & 48.1 (A$'$, $\nu_1$) & 50.8 (A$''$, $\nu_2$)  & 228.0 (A$'$, $\nu_3$) & 129.1     \\
    \hline
    \hline
    \end{tabular}
    \begin{tablenotes}
    \item Notes: $\nu_i$ and $\nu_j$ are two low-frequency out-of-plane bending modes, while $\nu_j$ is the most-off diagonal stretching mode.$\Delta \nu_{ij, k}^{(0)} = |\nu_k-\nu_i-\nu_j|$ is the unperturbed frequency spacing. The large splitting can mitigate the Fermi resonance coupling between the combination band $\nu_i\nu_j$ and the stretching mode $\nu_k$, as shown in Table \ref{tab:summary}.  All frequencies are calculated at the ground-state anharmonic-VPT level of theory.
    \end{tablenotes}
\label{tab:predict-molecules}
\end{table*}

\newpage
\begin{figure*}
    \centering
    \includegraphics[width = 1\textwidth]{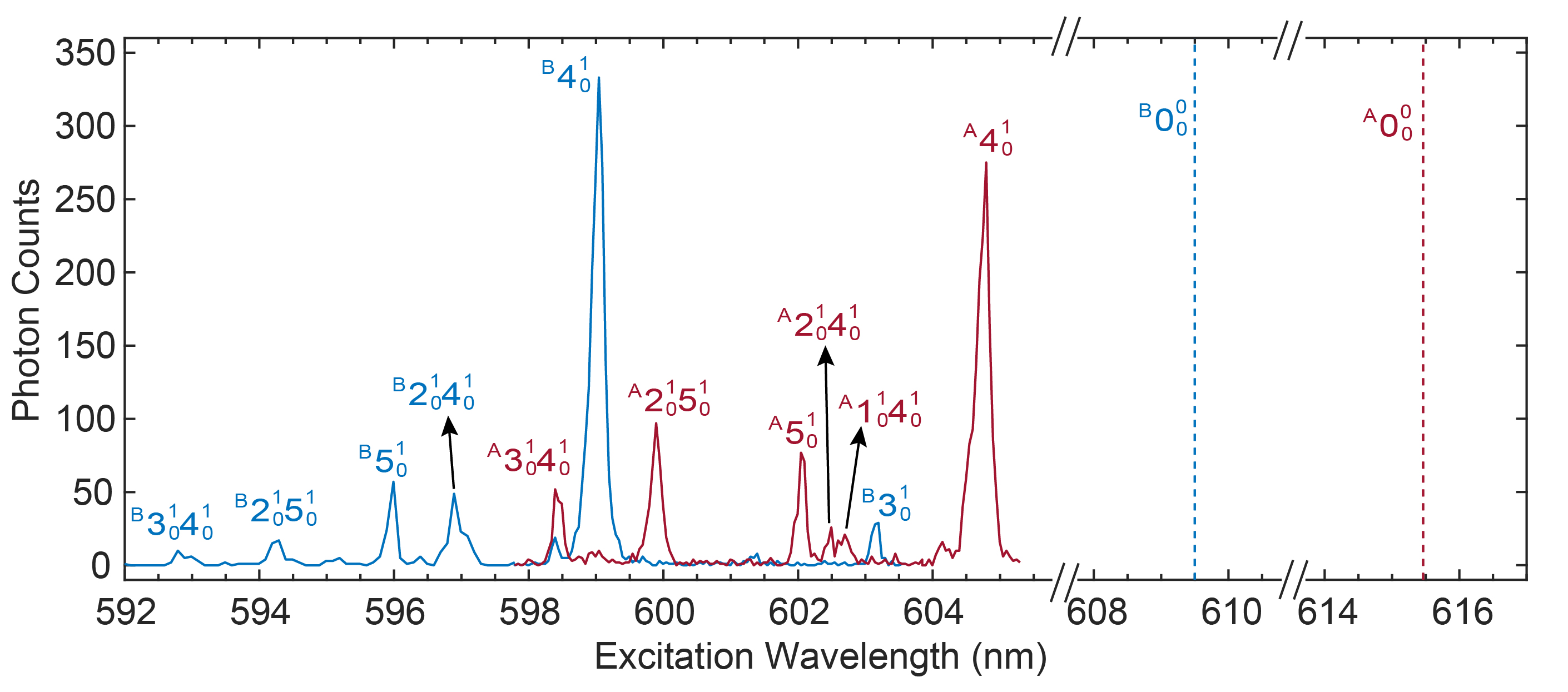}
    \caption{Excitation spectra for the excited states of CaOPh-4-F. The excitation wavelengths were scanned off-diagonally for $\widetilde A (v_n') \leftarrow \widetilde X (v'' = 0)$ (red trace) or $\widetilde B (v_n') \leftarrow \widetilde X (v'' = 0)$ (blue trace) transitions, while simultaneously monitoring the fluorescence photon counts at the diagonal 0-0 transitions. The two dashed lines indicate the excitation wavelengths corresponding to the respective 0-0 transitions. The assignments of all vibrational peaks, obtained by comparing with theoretical vibrational frequencies, are labeled and summarized in Table \ref{tab:vib-in-AB-Ca}.} 
    \label{fig:4F-excitaiton}
\end{figure*}

\begin{figure*}
    \centering
    \includegraphics[width = 0.95\textwidth]{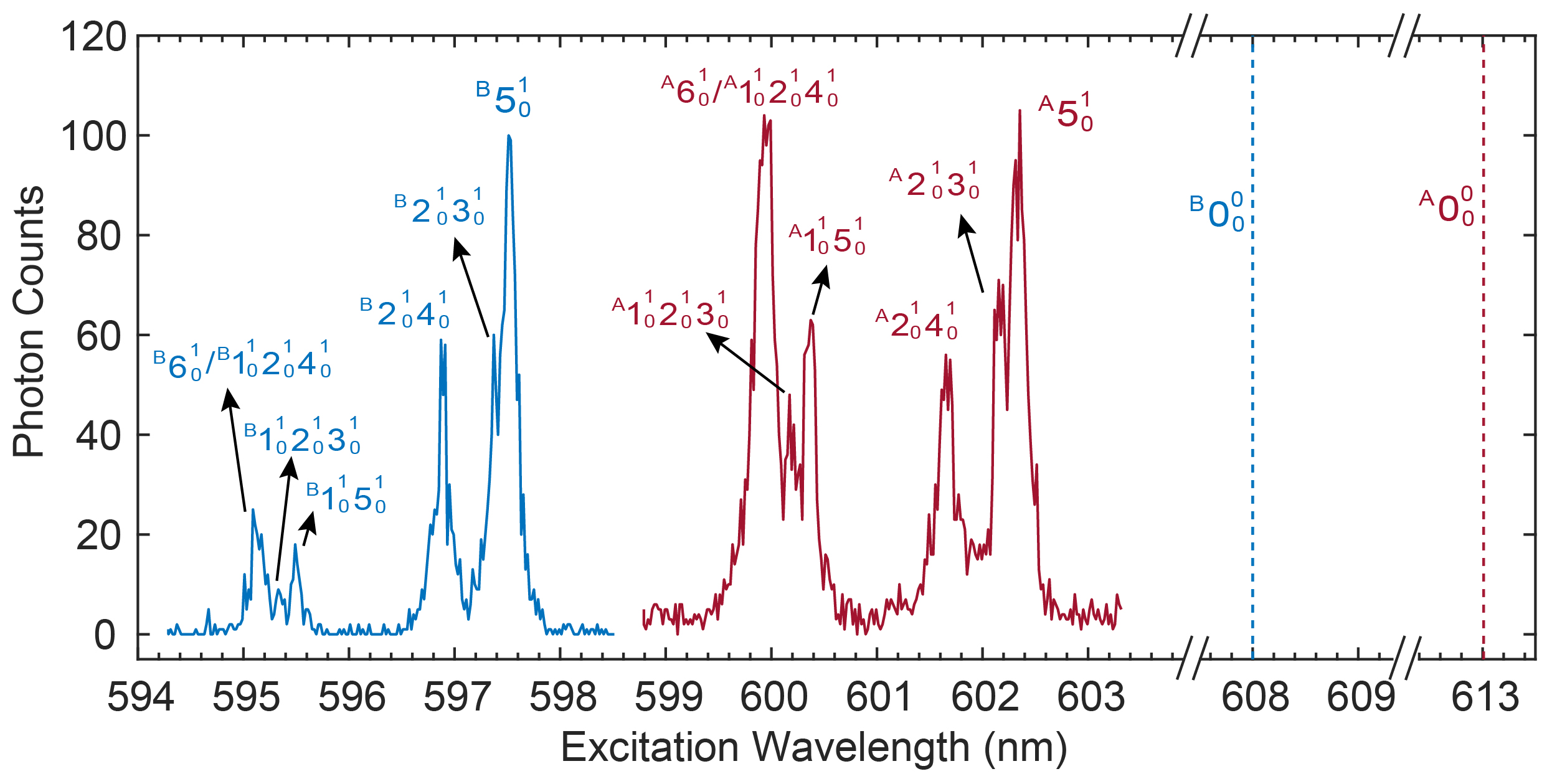}
    \caption{Excitation spectra of CaOPh-3-F. For $\widetilde A (v_n') \leftarrow \widetilde X (v'' = 0)$ (red trace) or $\widetilde B (v_n') \leftarrow \widetilde X (v'' = 0)$ (blue trace) transitions, the excitation wavelengths were scanned off-diagonally while simultaneously monitoring the fluorescence photon counts at the diagonal 0-0 transition. The two dashed lines indicate the excitation wavelengths of the respective 0-0 transitions. The assignments of all vibrational peaks, obtained by comparing with theoretical vibrational frequencies, are labeled and summarized in Table \ref{tab:vib-in-AB-Ca}.} 
    \label{fig:3F-excitation}
\end{figure*}

\begin{figure*}
    \centering
    \includegraphics[scale=0.5]{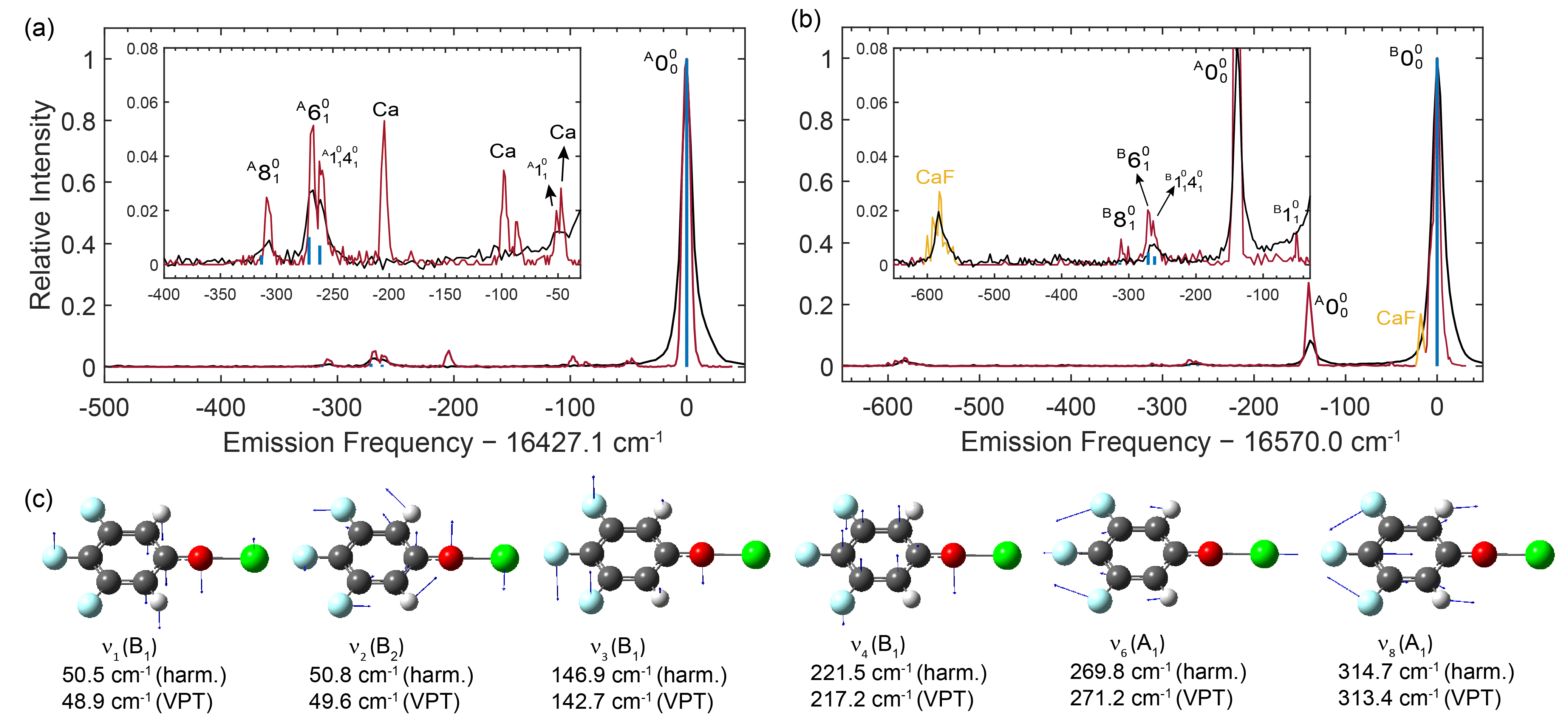}
    \caption{(a) and (b) Comparison of dispersed fluorescence spectra obtained for the $\widetilde A - \widetilde X$ and $\widetilde B - \widetilde X$ transitions of CaOPh-3,4,5-F$_3$ molecules, respectively, using two different measurements. The black traces are taken from ref.\cite{zhu2022caoph} with a spectral resolution of $\approx$ 0.5 nm. In contrast, the red traces represent new PMT measurements with an approved resolution of $\approx$ 0.20 nm. Notably, the red traces clearly exhibit the presence of splitting doublet peaks around $-270$ \cm. This is caused by the Fermi resonance between the stretching mode $\nu_6$ and a combination band $\nu_1\nu_4$ based on the vibrational frequencies and symmetries. The contamination of Ca atomic lines and CaF are observed (yellow traces). The blue vertical lines depict the calculated frequencies of the vibrational modes, while the height of the lines reflects their respective calculated strengths using the VPT method. The assignments and vibrational branching ratios of all resolved modes are summarized in Tables \ref{tab:DLIF-vbr} and \ref{tab:DLIF-vib}. (c) Vibrational displacements of six vibrational modes. Theoretical frequencies and symmetries for these modes are provided.} 
    \label{fig:345F-DLIF}
\end{figure*}

\begin{figure*}
    \centering
    \includegraphics[width = 0.95\textwidth]{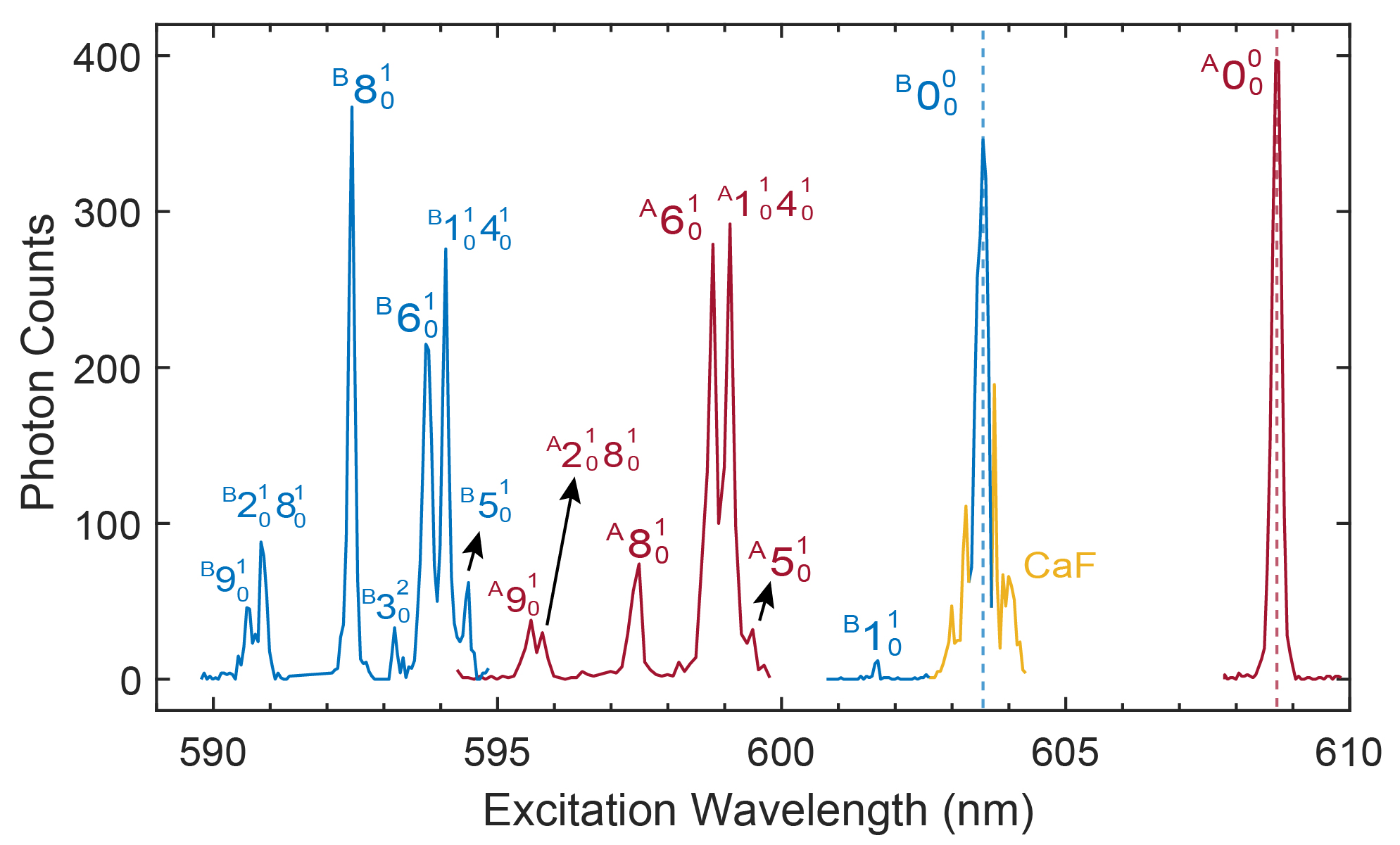}
    \caption{(a) Excitation spectra for the excited states of CaOPh-3,4,5-F$_3$. For $\widetilde A (v_n') \leftarrow \widetilde X (v'' = 0)$ (red traces) or $\widetilde B (v_n') \leftarrow \widetilde X (v'' = 0)$ (blue traces) transitions, the excitation wavelengths were scanned off-diagonally while simultaneously monitoring the fluorescence photon counts at the diagonal 0-0 transition. The two dashed lines indicate the excitation wavelengths of the respective 0-0 transitions. The $\widetilde B (v' = 0) \leftarrow \widetilde X (v'' = 0)$ transition is overlapped with the CaF transition (yellow traces). The assignments of all peaks, obtained by comparing with theoretical vibrational frequencies, are labeled and summarized in Table \ref{tab:vib-in-AB-Ca}.} 
    \label{fig:345F-excitation}
\end{figure*}

\begin{figure*}
    \centering
    \includegraphics[scale=0.95]{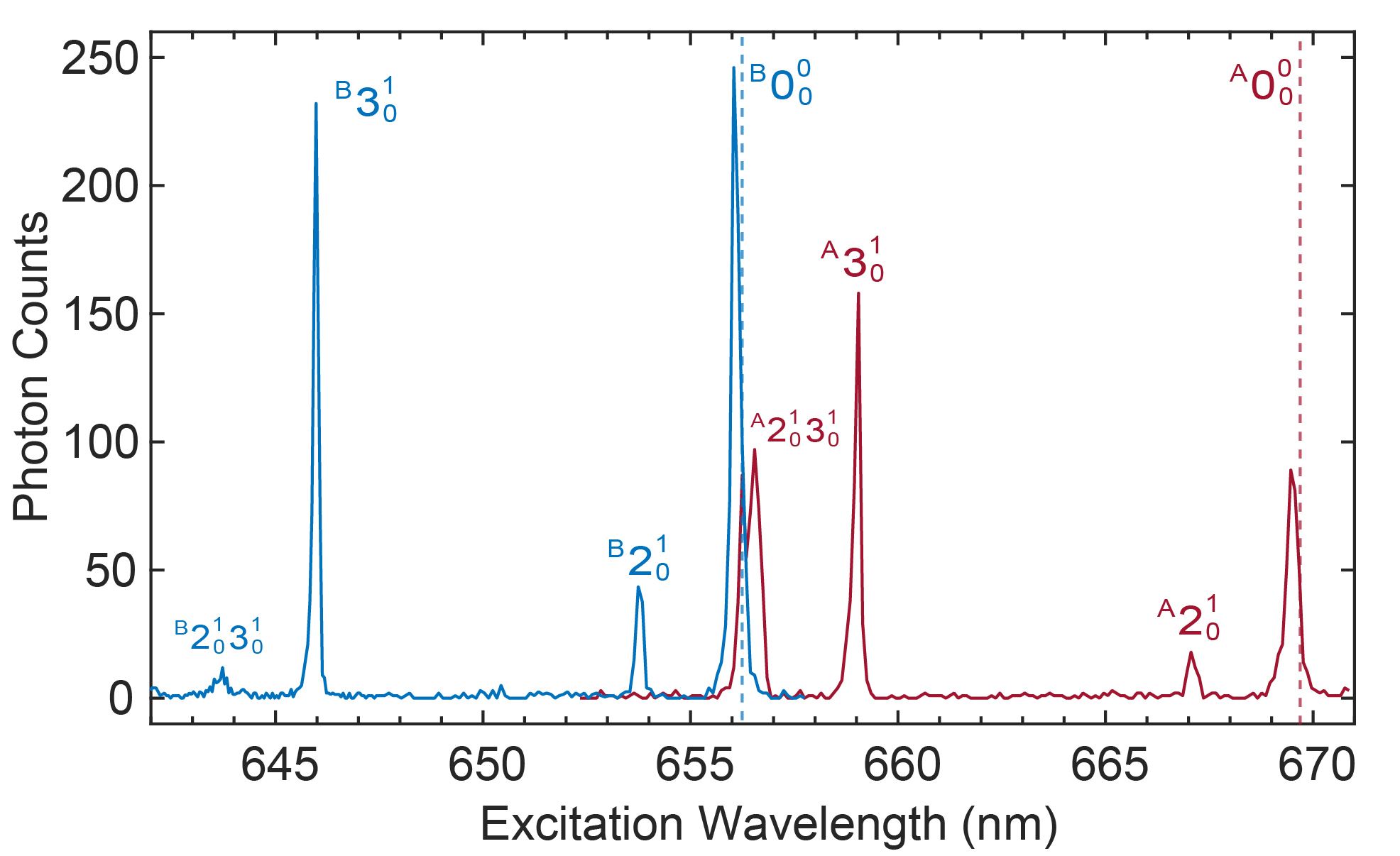}
    \caption{Excitation spectra for the excited states of SrOPh. For $\widetilde A (v_n') \leftarrow \widetilde X (v'' = 0)$ (red trace) or $\widetilde B (v_n') \leftarrow \widetilde X (v'' = 0)$ (blue trace) transitions, the excitation wavelengths were scanned off-diagonally while simultaneously monitoring the fluorescence photon counts at the diagonal 0-0 transition. The two dashed lines indicate the excitation wavelengths corresponding to the respective 0-0 transitions. The assignments of all vibrational peaks, obtained by comparing with theoretical vibrational frequencies, are labeled and summarized in Table \ref{tab:vib-in-AB-Sr}.} 
    \label{fig:sroph-excitation}
\end{figure*}

\begin{figure*}
    \centering
    \includegraphics[scale=0.65]{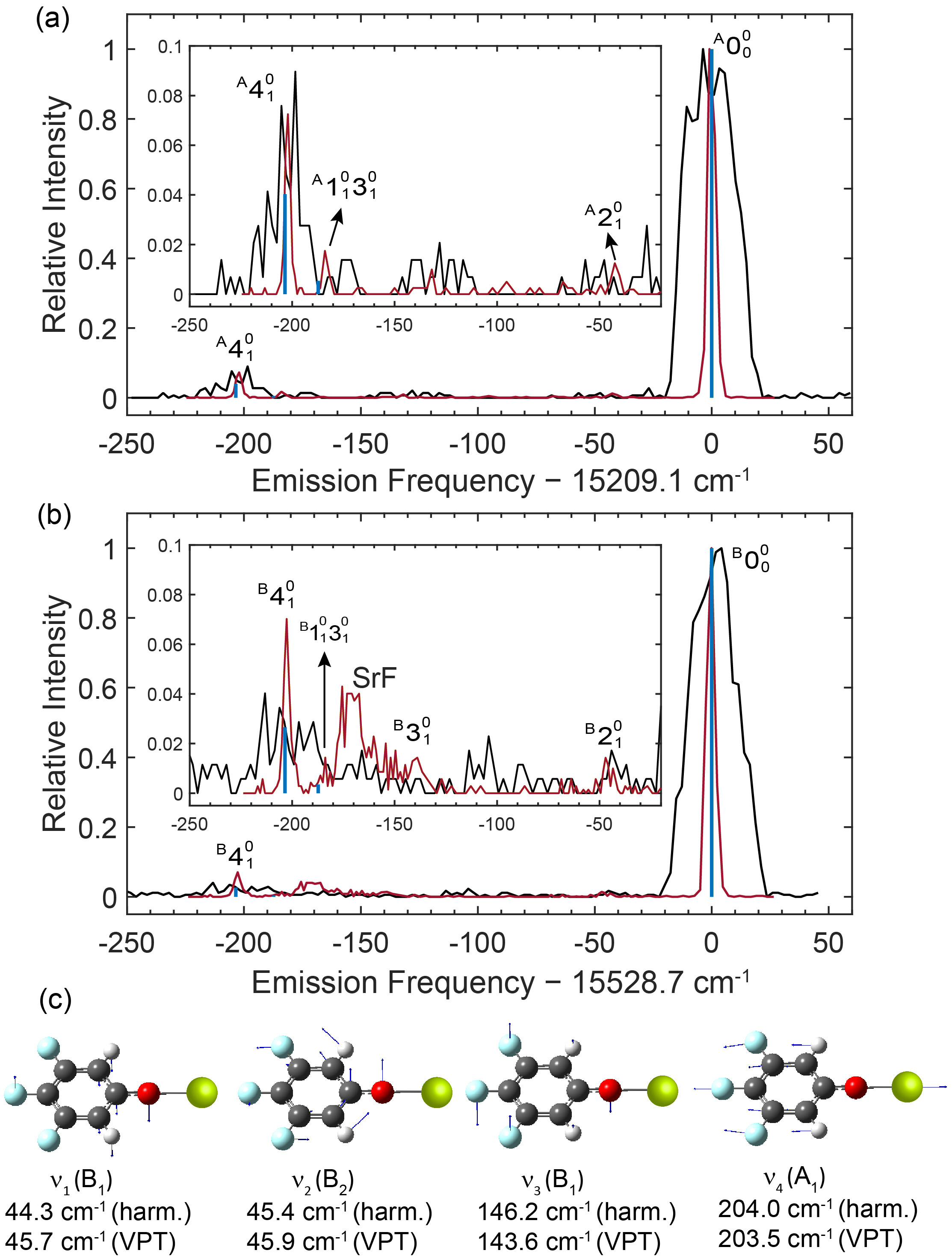}
    \caption{(a) and (b) Comparison of dispersed fluorescence spectra obtained for the $\widetilde A - \widetilde X$ and $\widetilde B - \widetilde X$ transitions of SrOPh-3,4,5-F$_3$ molecules, respectively. The black traces, reported previously \cite{lao2022sroph} have a low spectral resolution of $\approx$ 0.5 nm. In contrast, the red traces represent new measurements with a high resolution of $\approx$ 0.20 nm. A weak peak at around $-180$ \cm is assigned to the combination band of $\nu_1\nu_3$, which is due to the intensity borrowing from the Fermi resonance coupling with stretching mode $\nu_4$. The blue vertical lines depict the calculated frequencies of the vibrational modes, while the height of the lines reflects their respective calculated relative strengths using the VPT method. (c) Vibrational displacements of four related fundamental modes. Theoretical frequencies and symmetries for these modes are provided.} 
    \label{fig:sroph345f-dlif}
\end{figure*}

\begin{figure*}
    \centering
    \includegraphics[scale=0.95]{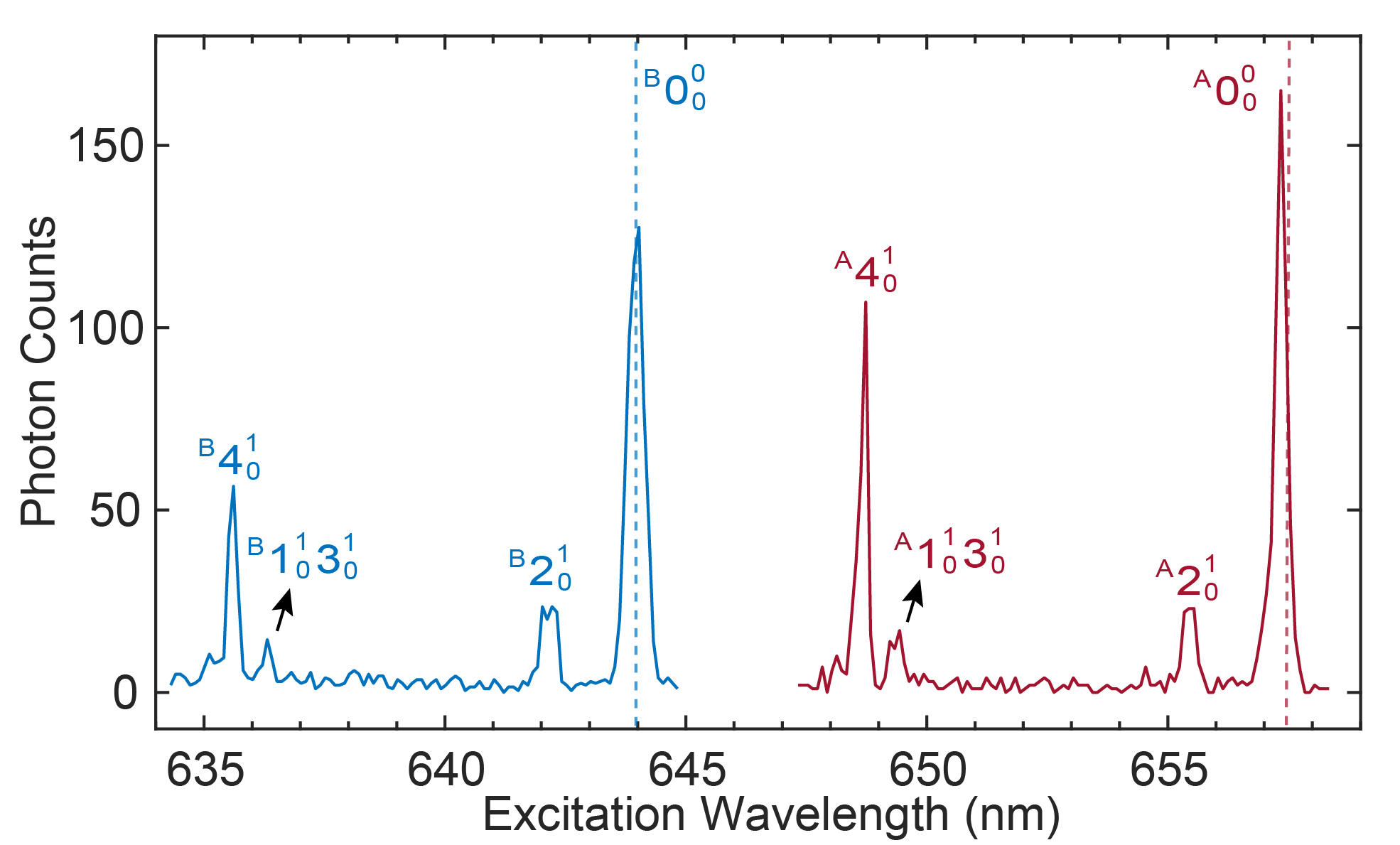}
    \caption{Excitation spectra for the excited states of SrOPh-3,4,5-F$_3$. For $\widetilde A (v_n') \leftarrow \widetilde X (v'' = 0)$ (red trace) or $\widetilde B (v_n') \leftarrow \widetilde X (v'' = 0)$ (blue trace) transitions, the excitation wavelengths were scanned off-diagonally while simultaneously monitoring the fluorescence photon counts at the diagonal 0-0 transition. The two dashed lines indicate the excitation wavelengths corresponding to the respective 0-0 transitions.  The assignments of all vibrational peaks, obtained by comparing with theoretical vibrational frequencies, are labeled and summarized in Table \ref{tab:vib-in-AB-Sr}.} 
    \label{fig:sroph345f-excitation}
\end{figure*}

\end{document}